\documentclass[pra,twocolumn,showpacs,preprintnumbers,
amsmath,superscriptaddress,amssymb]{revtex4}
\usepackage{graphicx} 
\usepackage{dcolumn} 
\usepackage{bm}
\usepackage{txfonts}
\usepackage{mathrsfs}
\usepackage{feynmf}
\usepackage{comment}
\usepackage{comment}

\begin{document}
\title{Quantum charge glasses of itinerant fermions with cavity-mediated long-range interactions} 

\author{Markus M\"uller}
\affiliation{The Abdus Salam International Center for Theoretical Physics, Strada Costiera 11, 34151 Trieste, Italy}
\author{Philipp Strack}
\email{pstrack@physics.harvard.edu}
\affiliation{Department of Physics, Harvard University, Cambridge MA 02138, USA}
\author{Subir Sachdev}
\affiliation{Department of Physics, Harvard University, Cambridge MA 02138, USA}
\affiliation{Instituut-Lorentz for Theoretical Physics, Universiteit Leiden, 
P.O. Box 9506, 2300 RA Leiden, The Netherlands}


\date{\today}

\begin{abstract}
We study models of itinerant spinless fermions
with random long-range interactions. We motivate such models from descriptions of fermionic atoms in multi-mode optical cavities.
The solution of an infinite-range model yields a metallic phase which has glassy charge dynamics, and a localized glass phase with suppressed density of states at low energies. 
We compare these phases to the conventional disordered Fermi liquid, and the insulating electron glass of semiconductors.
Prospects for the realization of such glassy phases in cold atom systems are discussed.
\end{abstract}

\maketitle

\section{Introduction}


There is much current interest in experiments with ultracold atoms and photons that provide clean realizations of 
models from condensed matter physics. A variant of the antiferromagnetic Ising model 
in one dimension, for example, has recently been ``quantum-simulated'' with bosons in optical lattices \cite{simon11} --an encouraging 
step toward quantum simulation of more general, strongly correlated quantum magnets in low dimensions. The hope is that these 
quantum optics experiments eventually reach the parameter regimes and accuracy necessary to allow for predictions that can overcome the limitations of conventional theoretical approaches for strongly interacting quantum many-body systems.

On top of that, the tunability of ultracold atoms allows one to explore new quantum many-body Hilbert 
spaces that have no direct condensed matter analog. In a series of remarkable experiments at ETH Zurich 
\cite{baumann10,baumann11,mottl12}, Baumann {\em et al.}, have begun the quantum-simulation of strongly-interacting quantum gases 
with genuine { \em long-range interactions} \cite{maschler05}. In these many-body cavity QED systems, an atomic ensemble (a thermal cloud \cite{domokos02,black03} or Bose-Einstein condensate) is loaded into an optical cavity containing quantized photon modes. Because the photons are massless, they mediate an interatomic interaction which does not decay as a function of distance between atoms and therefore 
 couples all the particles in the ensemble to one another. Exploiting this property, Baumann {\em et al.} \cite{baumann10,nagy10} 
found a 
mapping of their entangled atom-light system to the classic Dicke model describing $N$ two-level atoms uniformly coupled to a single quantized photon mode \cite{wang73,emary03,dimer07}. The superradiance transition of the $Z_2$ Dicke model spontaneously breaks an Ising symmetry and may be viewed as a realization of an Ising ferromagnet which is exactly solvable due 
to the infinite range of the photon-mediated ``spin-spin'' interaction.

Strack and Sachdev \cite{strack11} recently computed the phase diagram and spectral properties for atomic ``spins'' in multi-mode cavities assuming that the cavity mode functions and (fixed) positions of the atoms can be chosen so that the effective ``spin-spin'' interactions become {\em random} and {\em frustrated}. It was shown that a quantum phase transition in the 
universality class of the (solvable) infinite-range Ising quantum spin glass \cite{miller93,ye93} occurs, potentially enabling comparison 
between experiment and the theory of spin glasses. 
In a related paper \cite{gopa09,gopa10,gopa11}, Gopalakrishnan {\em et al.} provided experimental 
details and detection methods for the spin glass phase. 

In the present paper we explore quantum glassiness in the charge or density sector of {\em itinerant} fermionic atoms in 
multi-mode cavities. An important difference to the previous study~\cite{strack11} consists in the inclusion of hopping of atoms on the lattice; 
the resulting phase diagram now depends on the quantum statistics of the itinerant particles. Related bosonic versions (Bose Hubbard models coupled to 
cavity photons) were considered in Ref.~\onlinecite{silver10}, wherein a superradiant Mott insulating phase, displaying entanglement of the charge of the atoms with a cavity mode, was found. Optomechanical applications of degenerate fermions
in a cavity were considered previously in Ref.~\onlinecite{kanamoto10}. 

Our calculations provide evidence that multi-mode cavities with degenerate fermionic atoms can quantum-simulate 
various phases and properties of {\em infinite-range glasses} that share several properties with Efros-Shklovskii Coulomb glasses in the
quantum regime. 

For a single-mode cavity 
coupled to itinerant bosonic atoms \cite{baumann10}, the onset of superradiance is concomitant with translational symmetry breaking 
and the formation of a charge density wave with a period of half the cavity photon wavelength. The glassy ordering of fermionic atoms in multi-mode cavities can instead be understood as the formation of one out of many energetically low-lying amorphous charge density patterns in a random linear combination of cavity modes. Most of these amorphous density orderings are metastable 
and the ensuing slow relaxation dynamics should in principle be measurable as a response to 
probe laser fields. 

Glassy phases of fermions have long been predicted to exist in Coulomb frustrated semiconductors, so-called electron- or Coulomb glasses \cite{DaviesLeeRice}; in our case, the fermionic atoms play the role of the electrons in these earlier studies. In these electronic systems frustration results naturally from a competition between long range Coulomb repulsions and the random positions and energies of impurity sites. While the former favor a regular density pattern, such as in a Wigner crystal, the latter disrupt this order and provoke a non-crystalline density order. The competition of these two ingredients leads to many long-lived metastable configurations with very slow relaxation dynamics between them. This structure of phase space entails remarkable out-of-equilibrium properties 
such as memory effects and logarithmic relaxation persisting over many hours \cite{benchorin93, Grenet03,Popovic04} which 
are also present in quantum electron glasses approaching a metal insulator transition~\cite{ovadyahu07}. 
 
An important hallmark of such insulating electron glasses is the Efros-Shklovskii gap in the single particle density of states in the glassy phase, whose elementary charge excitations are strongly Anderson localized. Such a pseudogap is required by the stability 
of metastable states in the presence of unscreened long range interactions~\cite{efros75}. 
Remarkably, the amorphous ordering of the glass softens the hard gap --that would exist in a regularly ordered system-- 
just to the maximal extent which is still compatible with stability, leaving the glass in an interesting state of criticality~\cite{PankovDobrosavljevic,MuellerIoffe04}.
This criticality results in a widely distributed response to a local excitation and avalanches~\cite{Palassini}, features which occur also in mean field spin glasses~\cite{Pazmandi,Ledoussal}. 
This criticality survives in the presence of weak quantum fluctuations (non-zero tunneling amplitude between sites in electron glasses or spontaneous spin-flips induced by a transverse field in spin glasses).
It entails gapless collective excitations despite the absence of a broken continuous symmetry. Eventually the glass order melts at a 
critical value of the hopping or transverse field~\cite{ye93, sachdev95, dalidovich02,AndreanovMueller}.

Upon further increasing the quantum fluctuations, a common assumption is that the delocalization of the fermionic quasiparticles, {\em i.e.\/} the 
insulator-to-metal transition, co-incides with the disappearance of glassy dynamics, giving way to a disordered Fermi liquid. 
However, it is also possible that these two transitions are separated.
The Anderson delocalization of the fermionic quasiparticles may {\em precede} the melting of the glass, in which case we obtain a {\em metallic glass\/} 
with non-zero conductivity at zero temperature. Such a glassy state with metallic conduction was obtained in dynamical mean field theory
by Dobrosavljevic and collaborators \cite{pastor99,dobro03}.
Moreover, the fact that glassy phases may also exist in phases with good transport properties was recently shown in models of frustrated bosons~\cite{carleo09,tam10,yu11} where a 'superglass' phase with microscopically coexisting superfluid and glassy density order exists.
In the context of Coulomb frustrated systems in condensed matter (without disorder), an intermediate metallic phase with periodic, striped density order ("conducting crystal") was discussed by Spivak and Kivelson~\cite{Spivak}. 
 

\subsection{Overview of key results and outline of paper}

\begin{figure}
\vspace*{0mm}
\includegraphics*[width=89mm,angle=0]{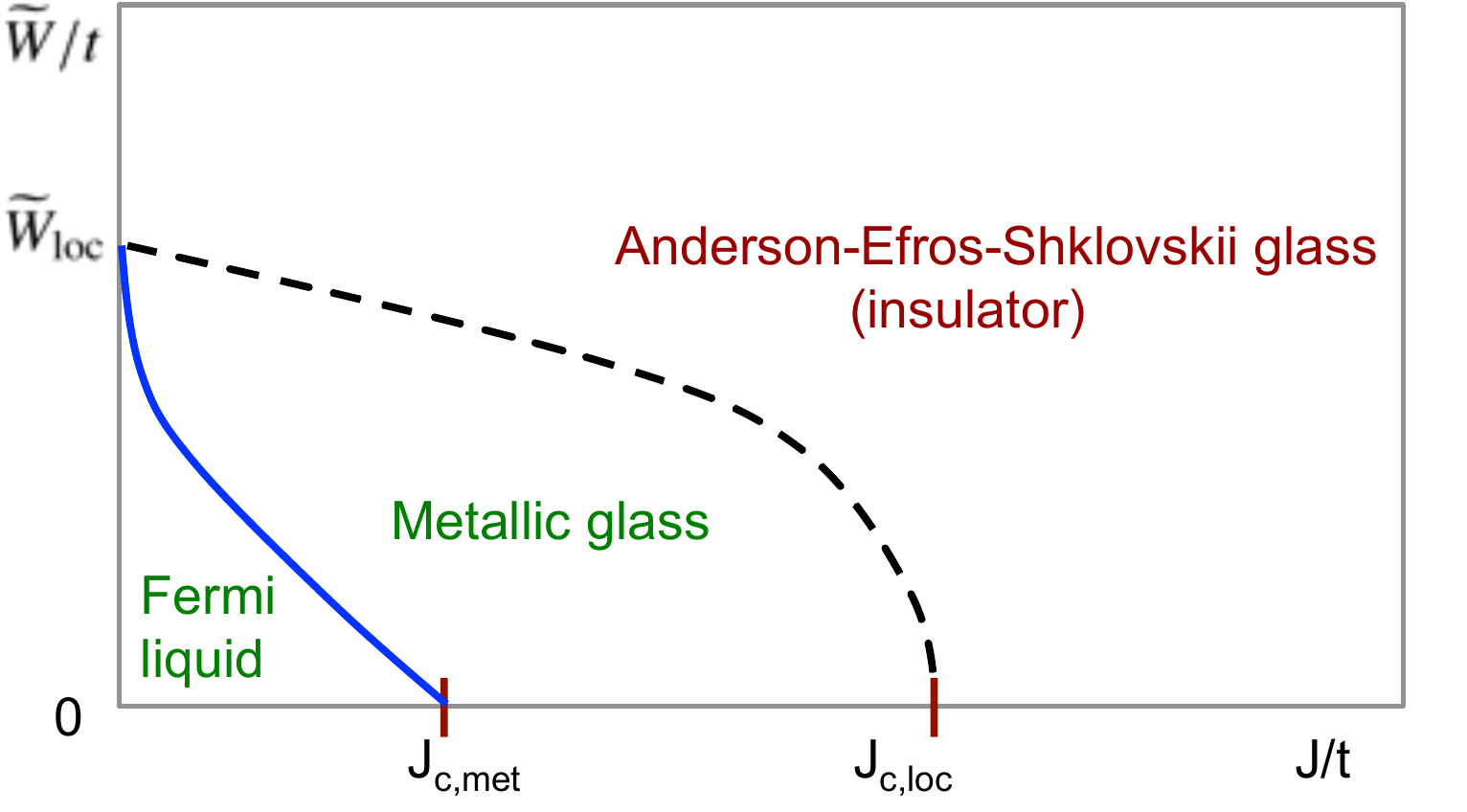}
\vspace*{0mm}
\caption{(Color online) Schematic phase diagram of model~(\ref{eq:fermi_hamiltonian_1}) for moderate densities 
in the vicinity of half-filling $n\approx1/2$ . $J^2$ is the variance of the long-range disorder $V_{ij}$, and $\widetilde{W}^2$ the variance of 
the effective onsite disorder, Eq.~(\ref{eq:onsite_variance}). The black, dashed line is the metal-insulator transition. The blue line is the glass transition. Quantitative computations in this paper 
are restricted to the $\tilde{W}=0$ axis, while the nature of the transition lines around the Anderson transition ($J=0$) are inferred from scaling considerations in Sec.~\ref{sec:fractal}. The localization transition from the metallic 
glass to the Anderson-Efros-Shklovskii glass and the signatures of this insulating glass phase are discussed in Sec.~\ref{sec:localized}. 
We describe the Fermi liquid to metallic glass 
transition and the properties of the metallic glass phase in Sec.~\ref{sec:itinerant}.}
\label{fig:phasediag_1}
\end{figure}

\begin{figure}
\vspace*{2mm}
\includegraphics*[width=89mm,angle=0]{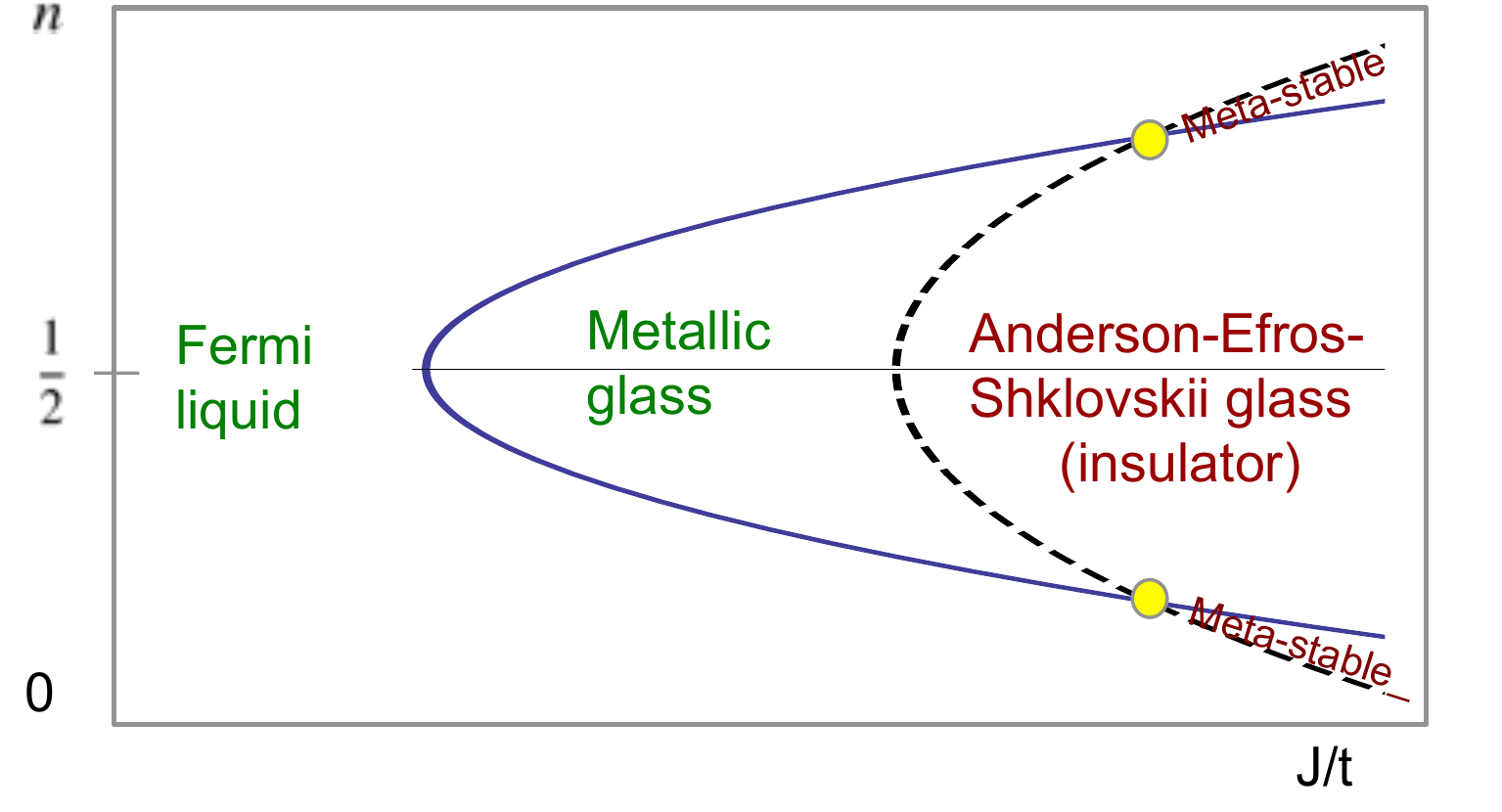}
\vspace*{-2mm}
\caption{(Color online) Schematic phase diagram of Eq.~(\ref{eq:fermi_hamiltonian_1}) for small effective onsite disorder 
($\widetilde{W}=0$) as a function of fermion density $n$. Below the crossing point (yellow dot) of the metal-insulator transition (black-dashed line) and the glass transition (blue), the transition 
becomes first order. The Fermi liquid and the metallic glass may still exist (as long-lived metastable states) to the right of the dashed line.}
\label{fig:density_phasediag}
\end{figure}

In the present paper, we argue that two types of glassy phases, a {\em metallic glass} and an {\em 
insulating Anderson-Efros-Shklovskii glass} also exist for fermions with random infinite-range interactions (see Fig.~\ref{fig:phasediag_1}). 
For low densities, we find that the metallic glass is avoided; instead, the liquid abruptly transitions to the localized glass state
(see Fig.~\ref{fig:density_phasediag}).

Our metallic glass state should not be confused with the metallic glass of metallurgy. In the latter materials, the glassy physics is entirely due to classical atoms freezing into out-of-equilibrium configurations, and the metalllic conduction is due to conduction electrons which move in the background of the frozen atoms. In contrast, in our system, the glassy dynamics and metallic conduction are due to the same fermionic degrees of freedom,
which are electrons in the condensed matter realizations, and fermionic atoms in quantum optics realizations.

In Sec.~\ref{sec:model}, starting from a Jaynes-Cummings type Hamiltonian for itinerant fermions coupled to cavity photons, 
we derive the fermionic model that we study in this paper:
\begin{align}
H=&-t\sum_{\langle i, j\rangle} \left(c^\dagger_{i}c_{j} + h.c.\right)
-\sum_{i=1}^N (\varepsilon_i-\mu) n_i
-\frac{1}{2}
\sum_{i,j=1}^N V_{ij} n_i n_j\;,
\label{eq:fermi_hamiltonian_1}
\end{align}
which contains a  {\em short-range} hopping term $t$, disordered, random onsite energies $\varepsilon_i$, and {\em long-range}, random density-density interactions $V_{ij}$ mediated by photons. 

In Fig.~\ref{fig:phasediag_1}, we show the phase diagram of this model at moderate density, $n=O(1)$, as a function of effective onsite disorder ($\widetilde{W}$ 
defined in Eq.~(\ref{eq:onsite_variance})) and photon-mediated interaction strength ($J$) in units of hopping $t$. 
For small effective onsite 
disorder $\widetilde{W}\gtrsim 0$, as the interaction is increased, the disordered Fermi liquid (FL) becomes unstable to the formation of an irregular, glassy density pattern, which depends on the interactions mediated by the random cavity modes. 
However, the irregular density waves do not gap the Fermi surface, but leave the fermions metallic, with a finite conductivity in the low temperature limit. The glassy charge density order is marginally stable, which leads to soft collective density excitations. The scattering of fermions from collective density modes leads to some 
non-Fermi liquid properties (such as finite-temperature transport) but the fermionic quasiparticles remain well-defined.
The quantum glass transition and properties of the metallic glass are analyzed with effective field theory methods in Sec.~\ref{sec:itinerant}.

Upon further reduction of the hopping, or increase of interactions, the random Hartree potential generated by the frozen density pattern starts to localize the quasiparticles and induces an Anderson insulator. The latter has strongly suppressed diffusion, which is expected to vanish at $T=0$. 
At this point, the metallic background vanishes, leaving behind an insulating charge glass with a spatially strongly fluctuating frozen-in density distribution. 
This phase and an estimate for the transition point 
$\left(J/t\right)_{c,\text{loc}}$ for a $3d$ cubic lattice is presented in Sec.~\ref{sec:localized}. We will show there that for low fermion densities, at equilibrium, the metallic glass is avoided and the Fermi liquid transitions discontinuously to a localized glass phase. Nevertheless, even at lower densities, the metallic glass may be experimentally observable, 
as it is expected to exist as a long-lived metastable phase, which eventually will 
nucleate the energetically more favorable localized glass. 

In Fig.~\ref{fig:phasediag_1}, the metallic glass ``strip'' is predicted to separate the AES glass from the Fermi liquid even close to the 
Anderson transition at $(J=0,\widetilde{W}_{\text{loc}})$. In that regime, as we explain in Sec.~\ref{sec:fractal}, 
the quantum glass transition is sensitive to the fractal nature 
of the fermionic wavefunctions close to the Anderson transition.

In Sec.~\ref{sec:conclu}, we conclude with a unifying discussion about the potential of many-body cavity QED as a quantum simulator
of long-range quantum glasses exhibiting freezing in different sectors
(spin, fermionic charge, bosonic charge). We also outline interesting open questions for future research. 

\section{Model}
\label{sec:model}

We consider spinless lattice fermions coupled to multiple cavity photon modes. The absorption of cavity photons (represented by canonical bosonic creation and annihilation operators $a^{\dagger}$, 
$a$) raises the internal state of the fermions from a ground state (represented by canonical fermion creation and annihilation 
operators $c^{\dagger}_{g}$, $c_{g}$) to an excited state ($c^{\dagger}_{e}$, $c_{e}$). In addition, a classically-treated pump laser 
coherently drives transitions between the ground state and the excited state. The Hamiltonian operator, 
\begin{align}
H[c_g^\dagger, c_g, a, c^\dagger_e, c_e]&=H_{\text{atom}}+H_{\text{photon}}+H_{\text{pump}},
\label{eq:micro_hamiltonian}
\end{align}
consists of three terms. The first one describes itinerant fermions on a d-dimensional (optical) lattice with $i=1,...,N$ sites with short-range, nearest-neighbor 
($\langle i, j\rangle$) hopping $t$ (taken to be the same for ground and excited state):
%
\begin{align}
H_{\text{atom}}=&-t\sum_{\langle i, j\rangle} \left(c^\dagger_{i,g}c_{j,g} + c^\dagger_{i,e}c_{j,e}+ h.c.\right)\nonumber\\
&+
\sum_{i=1}^{N}\left(\epsilon_{i,e}-\mu + \Delta\right)n_{i,e} +\sum_{i=1}^{N} 
\left(\epsilon_{i,g}-\mu\right) n_{i,g}\;.
\label{eq:hubbard}
\end{align}
Here $\mu$ is the chemical potential, $\epsilon_{i, g}$, $\epsilon_{i,e}$ are the single-particle energies of the ground and excited state, respectively, and 
$n_{i,g}=c^{\dagger}_{i,g}c_{i,g}$ are density operators for the atoms in the ground state, and analogously for the excited state.
Following Maschler {\em et al.} \cite{maschler08}, we write our equations in a frame rotating with the frequency of the pump laser 
and $\Delta$ describes the atom-pump detuning which can generally be chosen large $\Delta\gg t, \epsilon_{i,e}$.

We further have $M$ cavity photon modes with frequencies $\omega_{\ell}$ and spatially non-uniform atom-photon couplings $g_{i\ell}$ that generally depend on the atom's site $i$ and the characteristics of the cavity photon mode $\ell$. Finally, we have a pump term 
with amplitude $h_i$ that does not involve photon operators:
\begin{align}
H_{\text{photon}}&=\sum_{\ell=1}^{M}\omega_\ell a^{\dagger}_\ell a_\ell- \sum_{i=1}^N\sum_{\ell=1}^M g_{i\ell} 
\left(c^\dagger_{i,e} c_{i,g} a_\ell + c^{\dagger}_{i,g}
c_{i, e} a^{\dagger}_{\ell}\right)\;,\nonumber\\
H_{\text{pump}}&=-\sum_{i=1}^{N} h_i \left(c^{\dagger}_{i,e} c_{i,g}+ c^{\dagger}_{i,g} c_{i,e}\right)
\label{eq:pump}
\;.
\end{align}
The excited state can be adiabatically eliminated. In the aforementioned regime of large atom-pump detuning, one can ignore the dependence of the 
effective ground state interactions on the dynamics and spatial propagation of the excited state \cite{maschler08}. Therefore,  the $\epsilon_{i,e}$ do not appear in the parameters of the resulting effective Hamiltonian,
which reads: 
%
\begin{align}
H[c^{\dagger}_g,c_g,a]=&-t\sum_{\langle i, j\rangle} \left(c^\dagger_{i,g}c_{j,g} + h.c.\right)
+\sum_{i=1}^{N}\left(\epsilon_{i,g}-\mu+\frac{h_i^2}{\Delta}\right) n_{i,g}\nonumber\\
&+\sum_{\ell=1}^{M}\omega_\ell a^{\dagger}_\ell a_\ell
+\sum_{i=1}^N\sum_{\ell=1}^M\frac{g_{i\ell} h_{i}}{\Delta}
n_{i,g}\left(a_\ell + a^{\dagger}_\ell \right)\nonumber\\
&+\sum_{i=1}^N\sum_{\ell,m=1}^M\frac{g_{i\ell}g_{im} }{\Delta}n_{i,g} a^{\dagger}_{\ell}a_{m}\;.
\label{eq:ground_photons_hamiltonian}
\end{align}
%
We drop the subscript $g$ from now on. We include the expectation value of the term 
in the last line as a contribution to the single-particle energies, 
\begin{align}
\varepsilon_{i}=\sum_{\ell,m=1}^M\frac{g_{i\ell}g_{im} }{\Delta}\langle a^{\dagger}_{\ell}a_{m}\rangle +\epsilon_i+\frac{h_i^2}{\Delta}\;.
\label{eq:onsite_disorder}
\end{align}
In a final step, we integrate out the photons from Eq.~(\ref{eq:ground_photons_hamiltonian}) and get 
the expression:
\begin{align}
H\left[c^\dagger,c\right]=&-t\sum_{\langle i, j\rangle} \left(c^\dagger_{i}c_{j} + h.c.\right)
+\sum_{i=1}^N\left(\varepsilon_i-\mu\right)n_i
-
\frac{1}{2}
\sum_{i,j=1}^N V_{ij} n_i n_j\;.
\label{eq:fermi_hamiltonian}
\end{align}
The long-range density-density interaction written in a path integral representation
\begin{align}
V_{ij}(\Omega)=2\sum_{\ell=1}^M\frac{ g_{i\ell} g_{j\ell} h_i h_j}{\Delta^2} \frac{\omega_\ell}{\Omega^2+\omega_\ell^2}\;
\label{eq:photon_exchange}
\end{align}
makes the dependence on the bosonic Matsubara frequency $\Omega$ explicit~\footnote{Eq.~(\ref{eq:photon_exchange}) implies retarded interactions in Eq.~(\ref{eq:fermi_hamiltonian}), which should thus be written as an action, to be precise. However, in the low frequency limit discussed further below, we can approximate the interactions $V_{ij}$ as instantaneous.}.
The magnitude of $V_{ij}$ is proportional to the amplitude of the driving laser $h_{i}$ and can therefore be tuned flexibly. The sign and spatial dependence of $V_{ij}$ is determined by the choice of mode profiles of the cavity modes and pump lasers as well 
as the orientation of the lattice within the cavity.

In this paper, we are primarily interested in the case where the photon mode functions 
$g_{i\ell}g_{j\ell}$ in Eq.~(\ref{eq:photon_exchange}) can be realized as randomly varying in sign and magnitude 
in each disorder realization and with $M$ sufficiently large, 
we assume the $V_{ij}(\Omega)$ to be Gaussian-distributed with variance 
\begin{eqnarray}
\overline{\delta V_{i j} ( \Omega ) \delta V_{i' j'} (\Omega' )} &=& \left(\delta_{i i'} \delta_{j j'} + 
\delta_{j i'} \delta_{i j'} \right) V (\Omega, \Omega')/N.
\label{eq:V}
\end{eqnarray}
The overline represents a disorder average, and $\delta V_{i j}$ is the variation from the mean value. Such a mean value only shifts the chemical potential and can be dropped. Further, we assume in this paper that couplings between different sites are uncorrelated. 

We note that it should also be possible to generate random and frustrated interactions of longer range by using other means than random cavity modes. For example one might
employ a second fermion species to generate RKKY-type interactions 
among the primary fermion species. As in metallic spin glasses, such interactions decay as a power-law with distance 
and oscillate with periods of the Fermi wavelength of the second species, which induces frustration.


In the calculations below, the results only depend on the variance, which respects time-translation invariance
\begin{align}
V(\Omega,-\Omega)\equiv J^2(\Omega)\;.
\end{align}
To capture the main effects, as in Ref.~\onlinecite{strack11}, we may assume the simplified form:
\begin{align}
J(\Omega)=2 v^2 \omega_0/\left(\Omega^2+\omega_0^2\right)\;,
\label{eq:def_J}
\end{align}
where $\omega_0$ is a prototypical photon frequency representative of the spectral range of photons 
that mediate the inter-atomic interaction and $v$ the disorder strength.  For most of the paper we will concentrate on frequencies $\Omega\ll \omega_0$ and work with the static limit of the couplings
\begin{eqnarray}
\overline{\delta V_{i j} \delta V_{i' j'} } &=& \left(\delta_{i i'} \delta_{j j'} + 
\delta_{j i'} \delta_{i j'} \right) J^2/N, \nonumber\\
J&\equiv& J(0)= \frac{2v^2}{\omega_0}\,.
\label{eq:def_J0}
\end{eqnarray}
The photon contribution to the single-particle energies in Eq.~(\ref{eq:onsite_disorder}) and the pump term generate 
random local potentials for the fermions, which may additionally be superposed by a random lattice potential. We summarize all these effects by assuming random (Gaussian-distributed) onsite energies with independently tunable variance
\begin{align}
\overline{\delta\varepsilon_i \delta \varepsilon_j}=\delta_{ij}W^2\;.
\label{eq:W}
\end{align}
%


\section{Insulating Anderson-Efros-Shklovskii glass}
\label{sec:localized}

In the absence of hopping, $t=0$, the Hamiltonian (\ref{eq:fermi_hamiltonian}) reduces to the classical Sherrington-Kirkpatrick (SK) model~\cite{SK} of 
localized "spins" (describing presence or absence of a particle). This spins are subject to a 
random longitudinal field $\varepsilon_i$, and kept at fixed ``magnetization'' $M=2n-1$ where $n$ is the fermion density. The low temperature glass phase of this 
model is understood in great detail. 
As illustrated by the red graph in Fig.~\ref{fig:localfielddistribution}, a typical configuration of this glass exhibits a linear soft gap in the distribution
\begin{eqnarray}
P(\varphi) &=& \left\langle \frac{1}{N}\sum_{i=1}^N  \delta(\varphi-\varphi_i) \right\rangle
 \approx \alpha \frac{|\varphi|}{J^2}, \quad |\varphi |\lesssim J, 
 \label{pseudogap}
 \end{eqnarray}
of local Hartree fields 
\begin{eqnarray}
\varphi_i &\equiv& \frac{dH}{dn_i} =\varepsilon_i -\mu -\sum_{j\ne i} V_{ij}  n_j, 
\label{frozenfield}
 \end{eqnarray}
with coefficient $\alpha\approx 4\cdot0.31=1.24$~\cite{SommersDupont, Pankov06} and Gaussian decay for $|\varphi |\gg J$. The brackets 
$\langle ... \rangle $ stand for the thermodynamic average. As compared to the canonical SK model an extra factor of $4$  arises because we consider Ising degrees of freedom with magnitude $s^z_i\equiv n_i-1/2 =\pm 1/2$. Remarkably, the soft gap (\ref{pseudogap}) at low fields is universal, that is, {\em independent} of the strength of the random fields and the average magnetization (density)~\cite{markus07}. 
A similar soft gap, the Efros-Shklovskii Coulomb gap~\cite{efros75} is also known to exist in electron glasses with Coulomb interactions~\cite{pastor99,markus07}.

\subsection{Anderson-Efros-Shklovskii (AES) glass with quantum fluctuations/finite hopping ($t\neq0$)}
\label{subsec:aes_properties}

\begin{figure}
\vspace*{2mm}
\includegraphics*[width=86mm,angle=0]{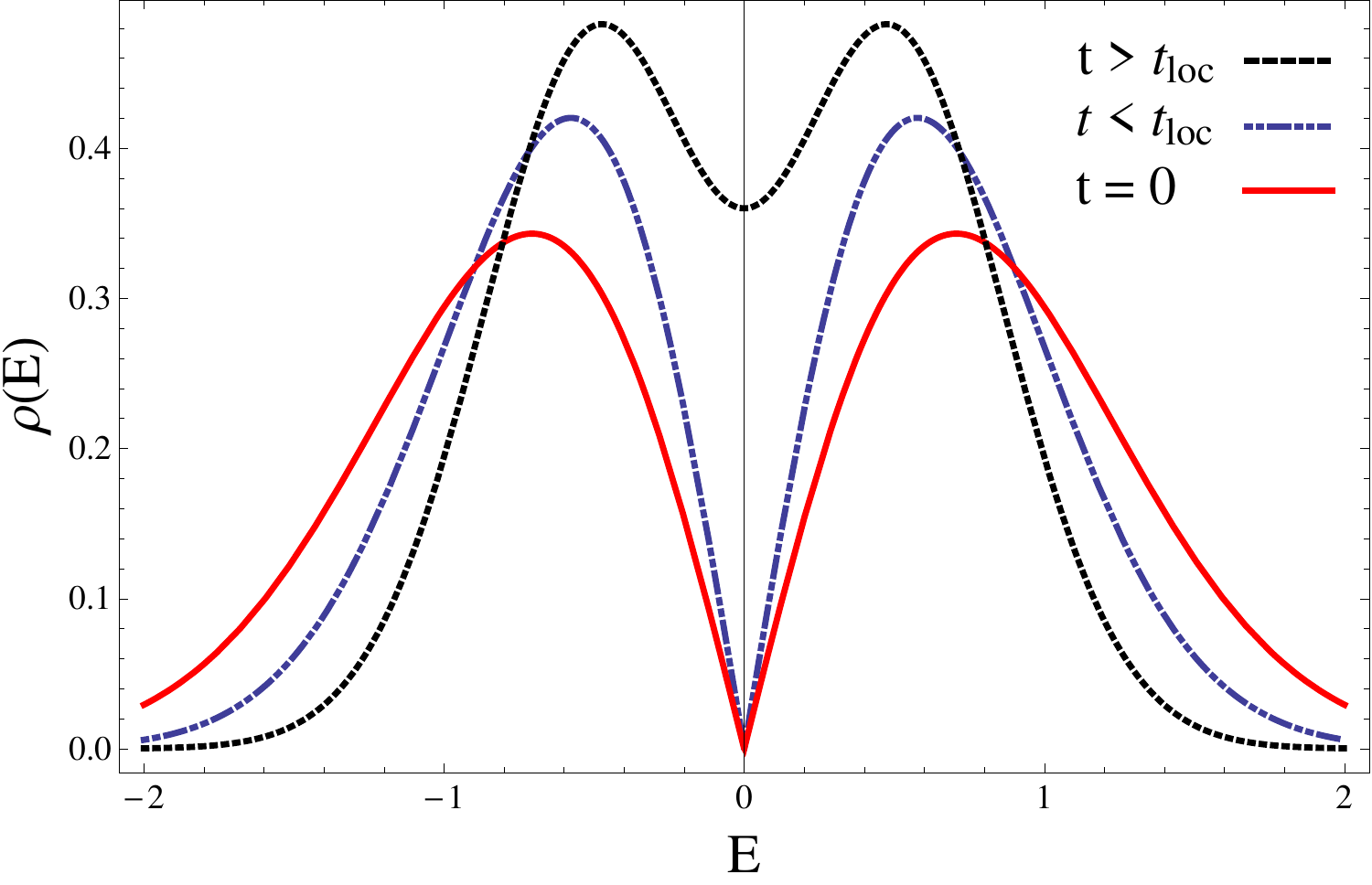}
\vspace*{-2mm}
\caption{(Color online) {Sketch of the evolution of the distribution of single particle energies $\rho(E)$ as the hopping $t$ increases. In the classical Efros-Shklovskii limit, $t=0$ (red solid curve), $\rho(E)$ coincides with the distribution of Hartree fields, which exhibits a linear pseudogap. In the localized AES glass at finite 
hopping (blue, dot-dashed curve) a pseudogap with vanishing density of states at the Fermi level $\rho(0)=0$ persists. A finite density of states at $E=0$ emerges when the quasiparticles delocalize and the system turns metallic. The dip in the density of states at low energy gradually weakens as $t$ increases.  }
}
\label{fig:localfielddistribution}
\end{figure}

Upon turning on quantum fluctuations via a finite hopping $t\neq0$, the fermions hop within the disorder potential of the above discussed Hartree fields. We refer to the eigenenergies of the resulting single particle problem as single particle excitations. In the limit $t\to 0$ 
the latter are completely localized and their energies coincide with the local Hartree fields. 
As long as the hopping stays below a critical value ($t< t_{\rm loc}$), the fermionic atoms remain Anderson-localized.  In that regime the single particle density of states 
in a given local minimum of the glass has to vanish at the Fermi level (down to energies of order $1/\sqrt{N}$, at $T=0$), otherwise the state would be unstable with respect to charge rearrangements. This follows from arguments analogous to those for quantum Coulomb glasses~\cite{efros75,Epperlein, LiPhilipps, Vignale}.
As a consequence, in a typical AES glass state, the linear compressibility (the analogue of zero-field-cooled susceptibility in spin glasses) vanishes, 
even though the full thermodynamic (field-cooled) compressibility is finite. Despite the vanishing linear compressibility, there is no hard charge gap, as there are charge excitations at any finite energy. 
The qualitative evolution of the distribution of single particle energies with increasing hopping is sketched in Fig.~\ref{fig:localfielddistribution}. 



%

We now explain the key differences of the AES glass to other previously discussed glass phases and describe its 
transport properties.

{\bf {\em (Quantum) electron glass:}} The glassy electrons in strongly doped semiconductors are localized mostly due to quenched potential disorder, whereas in the case of the AES glass the potential is mostly interaction-induced and thus self-generated. If $\widetilde W$ could be made to vanish, the 
localization would be entirely induced by the random interactions in a glassy frozen state. Both, the quantum electron glasses and 
the AES glass have linearly suppressed density of states at low energies. 

{\bf {\em Anderson or Fermi glass:}}  In these localized insulators, where quenched external disorder dominates, 
the quasiparticles are localized despite the presence of weak short range interactions. The bosonic analogues of such insulators are often referred to as {\em Bose glasses}~\cite{GiamarchiSchulz,Fisher}. Both fermionic and bosonic version of these glasses onsite-disorder dominated and do {\em not} exhibit glassiness in the sense of having many metastable configurations separated by high multi-particle barriers, except if strong interactions are present in addition to  the disorder potential. 
Another difference is that the strong interactions in the AES glass cause gapless collective modes, which are absent in less strongly correlated Fermi and Bose glass insulators. 

{\bf{\em Mott glass:}} In fermionic lattice systems at commensurate filling with strong nearest-neighbor interactions and disorder, this Mott glass arises as 
an intermediate phase which is incompressible due to a hard charge gap, but has a finite AC conductivity $\sigma(\omega)$ at any positive frequency \cite{ori99}. The main contribution to $\sigma(\omega)$ comes from local particle-hole excitations, which become gapless due to disorder. The AES glass is very different from this Mott glass, in that it has no hard charge gap, exists at any filling and does not rely on external potential disorder. 

{\bf{\em Ising spin glass in transverse field:}} 
If the occupation of sites $n_i=0,1$ is thought of as an Ising variable, the model (\ref{eq:fermi_hamiltonian}) looks similar to a mean field Ising spin glass in a transverse field $\Gamma$. The difference with the model considered in this paper consists mostly in the way in which quantum fluctuations act on top of the classical SK glass. This entails however a crucial difference of the respective phases to expect for weak long range interactions, as we will explain below.

The transverse field SK model is well known to undergo a quantum glass transition from a glass to a quantum disordered state at a critical value $\Gamma\sim J$.  The spins undergo glassy freezing when their interaction strength is bigger than the inverse polarizability of the softest two level systems. One can show that the thermodynamic glass transition persists in the presence of random longitudinal fields (the equivalent of the disorder potential $\varepsilon_i$ in Eq.~(\ref{eq:fermi_hamiltonian})), even though the latter spoils the Ising symmetry of the system. The glass transition then has the character of an Almeida-Thouless transition, like for mean field spin glasses in an external homogeneous or random field. The only symmetry which is broken at this transition is the replica symmetry, which is rigorously established to occur in models with infinite-range interactions, while it remains controversial in finite-dimensional, short range interacting systems. 

If the Ising states of a spin are thought of as two states of a localized fermion (two-level systems), the transverse field SK model describes interacting, but {\em fully localized} fermions.
It follows from the above discussion that such  systems have a non-glassy insulating phase. However, for our model a regime of localized but non-glassy fermions cannot exist. The reason is that there is no mechanism which gaps single fermion excitations away from zero energy, in contrast to the finite tunneling amplitude $\Gamma$ of two level systems which does gap the local excitations. Therefore, in the model considered here, the gapless low-energy fermionic states are always unstable to glass formation for any small infinite range interactions $V_{ij}$ (a similar argument was put forward in Ref.~\onlinecite{dobro03}). This is why the AES glass in Fig.~\ref{fig:phasediag_1} above $\widetilde{W}_{\text{loc}}$ extends all the way to the axis $J=0$. 

{\bf{\em Transport properties:}}
The main contribution to the finite temperature transport in the AES glass comes from inelastic scattering of the quasiparticles, with an inelastic rate which is expected to decrease as a power law of $T$. At low $T$, where this rate becomes smaller than the level spacing in the localization volume of the fermions, the charge transport is expected to proceed first by power law hopping, and at lowest $T$ by variable range hopping, assisted by inelastic processes among the fermions. This type of transport is drastically less efficient than in the metallic glass of Section \ref{sec:itinerant}. In the AES glass, the resistivity that should diverge with vanishing $T$. At the same time thermal transport may still be rather good due to collective density modes, which remain delocalized. 

It is tempting to speculate that, upon advancing the techniques of Ref.~\onlinecite{brantut12} to many-body cavity QED, the next generation of experiments could measure also the transport properties of neutral, glassy fermions.


\section{Delocalization transition out of the Anderson-Efros-Shklovskii glass at $\widetilde{W}\approx 0$}
%
 
As mentioned above, in the localized, insulating phase the Efros-Shklovskii stability argument assures a vanishing density of single particle states at the Fermi level.
However, the latter becomes finite when the quasiparticles delocalize at the insulator-to-metal transition, where diffusive transport behavior sets in. This takes place when the hopping becomes comparable to the typical potential difference between a site and its neighbors. A precise calculation of this transition would require the solution of the glass problem including quantum fluctuations, to obtain the distribution of effective onsite disorder, and the subsequent analysis of a delocalization transition of quasiparticles. However, an estimate for the critical value $\left(\frac{J}{t}\right)_{c,\text{loc}}$, 
below which the system behaves metallic, can be obtained as follows. 

\subsection{Moderate density: $n  \sim O(1)$}

Even if $\widetilde{W}$ of Eq.~(\ref{eq:onsite_variance}) is negligible, the soft gap in the Hartree potentials acts like a rather strong onsite disorder for the fermions. 
To discuss delocalization at the Fermi level we need to consider sites with Hartree potentials of the order of the hopping $\epsilon\lesssim t$. At such energies, the density of states in the soft gap can be roughly approximated by the constant density of states of a box-distributed onsite potential of width $W_{J}$ given by
\begin{eqnarray}
\label{Weff_loc}
W_{J}\approx  \frac{1}{P(\varphi \approx t)} \approx \frac{J^2}{\alpha t}\;.
 \end{eqnarray}
For a box distribution of onsite disorder, Anderson delocalization of quasiparticles on a cubic lattice in $d=3$ dimensions is known to occur at a critical disorder strength~\cite{Ohtsuki} 
\begin{eqnarray}
\left(\frac{W_c}{t}\right)_{3d}= Z_U\approx 16.54.
\end{eqnarray}
From this we may infer an estimate for the delocalization transition out of the insulating glass phase at
\begin{eqnarray}
\left(\frac{J}{t}\right)_{c,\text{loc}}\gtrsim \sqrt{\alpha Z_U} \approx 4.6. 
\end{eqnarray}
In this rough estimate we have neglected quantum fluctuations of the density order in the localized phase which are expected 
to increase the value of 
$\left({J/t}\right)_{c,\text{loc}}$ further. This is because they weaken the density inhomogeneity and thus the onsite disorder generated by the interactions.


\subsection{Low density: $n\ll 1$}
\label{subsubsec:low_dens}
At low fermion densities,  $n\ll 1$, the pseudogap is irrelevant for the delocalization 
transition because the gap is restricted to very small energies, far below the hopping strength needed for delocalization. 
The frozen fields (\ref{frozenfield}) have a typical magnitude $W_{\rm J}\sim \sqrt{n} J$. If this is the dominant contribution to disorder (i.e., for weak external disorder, $W\lesssim W_{\rm J}$), the delocalization instability arises in the glass phase when $t\sim W_{\rm J}$, i.e., when the interactions are reduced below the value 
\begin{eqnarray}
(J/t)_{\rm loc} \sim n^{-1/2}.
\label{Jtloc}
\end{eqnarray}
At larger external disorder, delocalization happens when $t\sim W$, independently of the interaction strength $J$.
However, the interaction strength (\ref{Jtloc}) is parametrically smaller than the instability $(J/t)_{\rm met} \sim n^{-5/6}$ of the Fermi liquid toward the metallic glass, which we will derive below in Eq.~(\ref{Jt_met}). This implies that at low densities, beyond the yellow, tri-critical points of Fig.~\ref{fig:density_phasediag}, the localized glass state jumps discontinuously into a non-glassy Fermi liquid via a first order transition.
The location of that transition can be estimated by considering the competition of kinetic energy cost $\sim t$ and potential energy gain $\sim\sqrt{n}J$ of transforming the Fermi liquid into a fully localized glass state. This yields 
\begin{eqnarray}
(J/t)_{\rm 1st}\sim n^{-1/2},
\end{eqnarray}
which scales in the same way with density as the delocalization instability of the localized phase, but is physically distinct from it. 
However, since the interactions are essentially infinite ranged, the system cannot simply nucleate a localized glass droplet locally, but has to undergo the localization transition more or less at once in the whole system. This suggests that the Fermi liquid phase is a very long lived metastable state, even far beyond the equilibrium transition point $(J/t)_{\rm 1st}$. 

\section{Metallic glass at $\widetilde{W}\approx 0$}
\label{sec:itinerant}

%
%

In this section, we develop an effective field theory approach for the Fermi liquid to metallic glass transition (the blue 
line for $\widetilde{W}\approx 0$ in Fig.~\ref{fig:phasediag_1}) and 
compute key properties of the metallic glass.
We can write the path integral pendant to 
Eq.~(\ref{eq:fermi_hamiltonian}) in terms of collective Hubbard-Stratonovich fields for charge fluctuations $\rho_i(\tau)$ 
and $N$ Lagrange multipliers $\alpha_i$ which enforce $\rho_i(\tau)=\bar{c}_i(\tau) c_i(\tau)$, 
where $\bar{c}$, $c$ are Grassmann variables representing the fermion fields and 
$T$ will denote temperature.  
With this field content, the partition function corresponding to the model of Eq.~(\ref{eq:fermi_hamiltonian}), 
$Z = \int D \alpha D \rho D \bar{c}Dc\,e^{-\big(\mathcal{S}_0+ 
\mathcal{S}_{\text{int}}
\big)}$, is determined by the action
\begin{align}
\nonumber\\
\mathcal{S}_0&=\int_0^{1/T} d\tau \sum_{i=1}^N
\bar{c}_i (\tau) \left(\partial_{\tau}-\mu\right) c_{i}(\tau)
\nonumber\\
&-\int_0^{1/T} d\tau \sum_{\langle i, j\rangle}
t\left( \bar{c}_i (\tau) c_j(\tau) + \bar{c}_j (\tau) c_i(\tau)\right)\;,\nonumber\\
\mathcal{S}_{\text{int}}&=-\frac{1}{2}\sum_{i,j=1}^N \int_0^{1/T} d\tau \int_0^{1/T} d\tau'
V_{ij}(\tau-\tau')
\rho_i(\tau)\rho_j(\tau')\nonumber\\
&+ \sum_{j=1}^N\int_0^{1/T} d \tau \,i\alpha_j \left(\bar{c}_j(\tau) c_j(\tau)-\rho_j(\tau)\right)
+
\rho_j(\tau)\varepsilon_j
\label{eq:fermi_bose_action}\;.
\end{align}
In order to streamline notation, we will use $\sum$ signs for all lattice site summations and integrations 
over imaginary time.

We now integrate out the fermions, performing a cumulant expansion in the terms $\propto \alpha$ to obtain an effective action for the density fields
$Z = \int D \alpha D \rho \,e^{-\mathcal{S}[\alpha,\rho]}$ with
\begin{align}
\mathcal{S}[\alpha,\rho]&=
-\frac{1}{2}\sum_{i,j,\tau,\tau'}
V_{ij}(\tau-\tau')
\rho_i(\tau)\rho_j(\tau')
+
\sum_{j,\tau} \rho_j(\tau)\varepsilon_j
\nonumber\\
&-
\sum_{j,\tau} \alpha_j(\tau)
\left(\langle n_j(\tau)\rangle_0 -\rho_j(\tau)\right)
\nonumber\\
&+
\frac{1}{2}\sum_{i,j,\tau,\tau'}
\alpha_i(\tau)\alpha_j(\tau')\langle n_i(\tau) n_j(\tau')\rangle_0
+...
\label{eq:mu_cumulants}\;,
\end{align}
where $\langle ... \rangle_0$ denotes the quantum and thermal average with respect to the non-disordered free fermion action $\mathcal{S}_0$. We introduce the auxiliary Hubbard-Stratonovich density fields $\rho_i$ as an intermediate step to make the physics more transparent and to make contact with our previous work \cite{strack11}. As in Ref.~\onlinecite{strack11}, we will later integrate them out again to obtain an action for the order parameter fields $Q_i^{ab}$. In App.~\ref{app:MillerHuse}, we present an alternative route to derive the final self-consistency equations  (\ref{eq:quadratic}-\ref{eq:q_saddle}), which offers insight into the nature of the approximations made below.

We proceed by integrating over $\alpha$, without keeping explicit track of the higher-order terms in $\alpha$ denoted by the dots 
in Eq.~(\ref{eq:mu_cumulants}). 
Below we drop those corrections. However, the alternative approach in App.~\ref{app:MillerHuse} can in principle resum them, at the expense of replacing the bare density-density correlator Eq.~(\ref{Pi_0}) with the full proper ``polarizability'' \cite{miller93}. By dropping the interaction corrections to the latter, we operate at a level equivalent to the approximation used by  Miller and Huse for the closely related infinite-range quantum spin glass problem~\cite{miller93}. Those authors obtained a good estimate for the quantum glass transition point when 
compared with more elaborate studies \cite{grempel_added, ishii_added} .

It is convenient to express the resulting action in terms of local fluctuations in the onsite density $\delta \rho_i(\tau)=
\rho_i(\tau)-\langle n_i(\tau)\rangle_0\equiv \rho_i(\tau)- n$: 
%
\begin{align} 
\mathcal{S}[\rho]&=
\frac{1}{2}\sum_{i,j,\tau,\tau'} 
\delta\rho_i(\tau)\left[-\Pi^{(0)}\right]^{-1}_{\tau,\tau',i,j}\delta \rho_j(\tau')
\label{Srho2} \\
&-\frac{1}{2}\sum_{i,j,\tau,\tau'}\delta \rho_i(\tau)V_{ij}(\tau-\tau')\delta\rho_j(\tau')
+
\sum_{\tau,i}\delta \rho_i(\tau) \tilde{\varepsilon}_i+...\;,\nonumber
\end{align}
with the bare density-density correlator denoted by
\begin{align}
\Pi^{(0)}_{i,j,\tau,\tau'}= \langle n_i(\tau) n_j(\tau')\rangle_0\;.
\label{Pi_0}
\end{align}
The effective onsite disorder consists of two terms
\begin{align} 
\tilde{\varepsilon}_i=\varepsilon_i- \sum_{j,\tau'}\langle n_j(\tau')\rangle_0 V_{ij}(\tau-\tau')\;.
\label{effdis}
\end{align}
Focusing on low frequencies we neglect the retardation, and find the disorder variance from Eqs.~(\ref{eq:V}, \ref{eq:W}) as:
\begin{align}
\overline{\delta \tilde{\varepsilon}_{i}\, \delta\tilde{\varepsilon}_{j}}
&=\delta_{ij}\left(W^2+  J^2(0) \overline{\langle n_k\rangle^2} \right) \equiv
\delta_{ij} \widetilde{W}^2\;.
\label{eq:onsite_variance}
\end{align}
The dots in (\ref{Srho2}) stand for interactions between more than two $\delta\rho_i$.

We proceed by employing standard replica methods \cite{binder86} to average over the disorder configurations of 
the $V_{ij}(\Omega)$ and $\varepsilon_i$. The resulting 
inter-replica interaction term $\sim\delta\rho^4$ proportional to the variance of $V_{ij}$  is non-local in both, imaginary time and position space. We decouple this terms with 
an inter-replica matrix-valued Hubbard-Stratonovich field, that depends on two frequencies,
$Q^{ab}_{i}(\Omega,\Omega')\leftrightarrow \delta \rho_i^a(\Omega)\delta\rho^b_i(\Omega')$, which is bi-local in imaginary time, but local 
in position space. Finally, we write the $n$-times replicated, disorder-averaged partition function 
$\overline{Z^n}=  \int D Q D \rho \,e^{-\bar{\mathcal{S}}}$ with the action
\begin{align}
\bar{\mathcal{S}}=&
\frac{1}{2}\sum_{i,j,\tau,\tau',a} 
\delta\rho^a_i(\tau)\left[-\Pi^{(0)}\right]^{-1}_{\tau,\tau',i,j}\delta \rho^a_j(\tau')
-\sum_{i,\tau,\tau',a,b} \frac{\widetilde{W}^2}{2} \delta\rho^a_i(\tau) \delta\rho^b_i(\tau')
\nonumber\\
&+ 
T^2\sum_{i,j,\Omega,\Omega',a,b}
\frac{V(\Omega,\Omega')}{N}\Bigg[
\frac{1}{4}
Q_i^{ab}(\Omega,\Omega')
Q_j^{ab}(-\Omega,-\Omega')
-
\nonumber\\
&
\frac{1}{2} 
Q_i^{ab}(-\Omega,-\Omega')
\delta \rho^a_j(\Omega) 
\delta \rho^b_j(\Omega')
\Bigg]\label{eq:action_Q_lambda_rho}\;.
\end{align}
In the end we will take the replica limit $n\to 0$ to extract quenched averages. The dots stand for further replica-diagonal interactions between several $\delta \rho_i$, which are generated by the higher order cumulants in Eq.~(\ref{eq:mu_cumulants}). 

Incorporating local disorder into the polarizablity, 
which replaces $\Pi^{(0)}_{i,j,\tau,\tau'}$ in a complete solution will actually generate replica off-diagonal terms in $\Pi$ that self-consistently depend on $\widetilde{W}$ and $J^2(0) \overline{\langle n\rangle^2}$. This significantly complicates the analysis. 
As announced above, we restrict here formally to the $\widetilde{W}=0$ slice of Fig.~\ref{fig:phasediag_1}, and defer the quantitative analysis of onsite-disorder effects to future work~\footnote{At the glass transition, for $W=0$, the effective disorder $\widetilde{W}$ is not very strong as compared to the hopping. Therefore, we hope that our estimates for the instability of the Fermi liquid phase to the metallic glass, 
$(J/t)_{\rm met}$, are reasonable even for small finite $\widetilde{W}$. The neglect of the effective disorder probably overestimates the resilience of the Fermi liquid to form a density pattern, and hence is expected to yield an overestimate for  $(J/t)_{\rm met}$.}.


\subsection{Saddle-point free energy  ($\widetilde{W}=0$)}
\label{SPfreeenrgy}

Without any truncation in the cumulant expansion (\ref{eq:mu_cumulants}), the functional integration would yield results for the correlators of $\delta \rho$ that automatically 
obey the Pauli principle. However, after a truncation the latter might be violated. In order to proceed with 
a saddle-point analysis, we introduce into the action a global Lagrange multiplier $\lambda^a$ conjugate to density fluctuations, enforcing that 
correlation functions still obey the Pauli principle on average. Namely, $\lambda^a$ is 
determined self-consistently  so that the equality
\begin{align}
\sum_{i=1}^N\langle\delta \rho^a_i(\tau) \delta\rho^a_i(\tau)\rangle=  Nn (1 - n) 
\end{align}
is satisfied. The lefthand side can be viewed as the variance of the binomial probability distribution 
of $N$ lattice sites being either occupied ($\rho_i = 1$) with probability $n$ or empty ($\rho_i = 0$) 
with probability $1-n$. We recall that $n$ is the average density. $n=0$ corresponds to an empty band, $n=1/2$ to half-filling, 
and at $n=1$ the band is filled. 

  
This global constraint is analogous to the ``spherical'' approximation in spin glasses. It can be shown that relaxing the (exact) local constraints to a global constraint does not affect the quantum critical behavior~\cite{kosterlitz76,almeida78,ye93,sengupta95}.

Noticing that the action in Eq.~(\ref{eq:action_Q_lambda_rho}) is quadratic in the $\rho$-variable, 
we can integrate it out exactly. Note that the bare density correlator $\Pi^{(0)}_{\tau,\tau',i,j}$  
only depends on the differences in time ($|\tau-\tau'|$) and space ($|i-j|$).
The resulting action has a prefactor $N$, the number of atoms, and thus, a saddle-point evaluation becomes exact. 
The free energy
\begin{align}
\overline{\mathcal{F}}=-\lim_{n\rightarrow0}\lim_{N\rightarrow\infty} \frac{T}{Nn}\ln \overline{Z^n}\;,
\label{eq:def_free_energy}
\end{align}
becomes a functional of the order parameter field 
\begin{align}
Q^{ab}(\tau,\tau')\equiv \frac{1}{N}\sum_iQ_i^{ab}(\tau,\tau') = \frac{1}{N}\sum_i\langle \delta\rho_i^a(\tau)\delta \rho^b_i (\tau')\rangle
\label{Qab}
\end{align}
and the Lagrange multipliers $\lambda^a$. Assuming that on the saddle point the latter take a replica-symmetric value $\lambda$, the functional takes the form $\overline{\mathcal{F}}=  \lim_{n\rightarrow0}\frac{1}{n} \overline{\mathcal{F}}_n$ with 
\begin{align}
\overline{\mathcal{F}}_n(Q^{ab},\lambda) =& 
\frac{T}{4} \sum_{a,b,\Omega} J^2(\Omega) \big | Q^{ab}(-\Omega,\Omega) \big |^2
-\lambda \left(1-n\right)n
\nonumber\\
&+ \frac{T}{2} \sum_{\omega,\mathbf{q}}
\text{tr}_{\text{ab}} \ln
\Bigg[
\left(
\left[-\Pi^{(0)}(\Omega,\mathbf{q})\right]^{-1}
+ 2\lambda\right)\delta^{ab}
\nonumber\\
&
\quad\quad\quad\quad\quad -J^2(\Omega) Q^{ab}(-\Omega,\Omega)
\Bigg]\;,
\label{eq:general_free_energy}
\end{align}
where $\text{tr}_{\text{ab}}$ denotes the trace over replica indices and the dots stand for the neglected terms mentioned above. 
We have imposed time-translational invariance of $Q^{ab}$: 
$$Q^{ab}(\Omega,\Omega')\rightarrow Q^{ab}(\Omega,\Omega') \delta_{\Omega+\Omega',0}/T. $$

%
%

As long as we are primarily interested in the critical behavior and the properties of a single typical state close to the glass transition, we may proceed with a replica symmetric calculation, even though the replica-symmetric saddle point is strictly speaking unstable towards the breaking of replica symmetry. The latter signals the emergence of many pure states in the glass phase, the breakdown of full thermalization, i.e., ergodicity breaking and the associated out-of-equilibrium phenomena at long time scales.  As we show in App.~\ref{app:MillerHuse}, the glass 
instability conditions obtained below (Eq.~\ref{eq:quadratic},\ref{eq:instability}) indeed signal the instability of a replica symmetric saddle-point 
toward replica symmetry breaking. 

For the $Q^{ab}$ fields, the following ansatz is natural:
\begin{align}
Q^{ab}(\Omega,-\Omega)&=
D(\Omega)\delta^{ab}+\frac{\delta_{\Omega,0}}{T}q_{\text{EA}}\;.
\label{eq:ansaetze}
\end{align}
Here, the Edwards-Anderson order parameter $q_{\text{EA}}$ shows up both in diagonal and off-diagonal entries of $Q^{ab}$. 
This ansatz in terms of $q_{\text{EA}}$ and the (site- and disorder-averaged) dynamic density response $D(\Omega)$ is the most general one, respecting replica symmetry and time-translational invariance. 

A non-zero value of $q_{\text{EA}}$ signals a frozen-in density distribution of the atoms:
\begin{align}
q_{\text{EA}}&=\lim_{t\rightarrow \infty} \frac{1}{N}\sum_\ell
\overline{
\big\langle
\delta \rho_\ell(t) \delta \rho_\ell(0)
\big\rangle
}\;.
\nonumber
\end{align}
In the glass phase, the spatially non-uniform on-site densities differ randomly from their average value (depending on the state into which the glass freezes) and retain that value for infinitely long times. As in any glass, these frozen density patterns are expected to depend sensitively on the details of the quench protocol or the preparation history in general. 
Note however, that only in the case of vanishing effective disorder, $\tilde W=0$, $q_{\rm EA}$ serves as an order parameter for the glass 
transition. In the more realistic case of finite onsite disorder, $\tilde W\neq 0$, the system still exhibits a thermodynamic glass transition, as do mean field spin glasses in random fields, but the only symmetry to be broken in that case is the replica symmetry, since density inhomogeneities already exist in the Fermi liquid phase (cf. App.~\ref{app:MillerHuse}). 

The average dynamic density response,
\begin{align}
D(\Omega)=\frac{1}{N}\sum_\ell \overline{
\big\langle
\delta\rho_\ell(\Omega) \delta\rho_\ell(-\Omega)
\big\rangle}\;,
\end{align}
characterizes the response of the fermions to local, time-periodic modulations of the density. 
In terms of the parametrization (\ref{eq:ansaetze}), the free energy~(\ref{eq:general_free_energy}) obtains as 
\begin{align}
\overline{\mathcal{F}} = &
\frac{T}{4} \sum_\Omega J^2(\Omega)  | D ( \Omega ) |^2  + 
\frac{1}{2} J^2(0) D(0) q_{\text{EA}} -\lambda \left(1-n\right) n 
\nonumber\\
&+ 
\frac{T}{2} \sum_\Omega \int \frac{d^d \mathbf{q}}{\left(2\pi\right)^d}
 \ln \Bigg[\left[-\Pi^{(0)}(\Omega,\mathbf{q})\right]^{-1}
 + 2\lambda
 - J^2(\Omega)D(\Omega) \Bigg]
\nonumber\\
 & - \frac{1}{2}\int \frac{d^d \mathbf{q}}{\left(2\pi\right)^d}
 \left[\frac{J^2(0) q_{\text{EA}}}{ \left[-\Pi^{(0)}(\Omega,\mathbf{q})\right]^{-1} + 2\lambda - J^2(0) D(0) }\right]\;.
\label{eq:free_energy}
\end{align}
This free energy depends on the microscopic parameters of the original fermionic theory via the 
bare density response
\begin{align}
\langle n_i(\tau)n_j(\tau')\rangle_0 &=T\sum_{\Omega}
\int \frac{d^d \mathbf{q}}{\left(2\pi\right)^d}
e^{-i\Omega(\tau-\tau')+i\mathbf{q} \left(\mathbf{x}_i-\mathbf{x}_j\right)}
\Pi^{(0)}(\Omega,\mathbf{q})\;,
\end{align}
where $\Pi^{(0)}$ is the real part of the particle-hole bubble, i.e. the convolution of two bare fermion propagators: 
\begin{align}
\Pi^{(0)}(\Omega,\mathbf{q})
&=\text{Re}\Bigg[-T\sum_{\omega}\int\frac{ d^d \mathbf{k}}{\left(2\pi\right)^d} G_0(\omega+\Omega,\mathbf{k}+\mathbf{q}) 
G_0(\omega,\mathbf{k}) \Bigg]
\nonumber\\
&
=\text{Re}\Bigg[\int\frac{ d^d \mathbf{k}}{\left(2\pi\right)^d}
\frac{f(\xi_{ {\bf k} })- f(\xi_{{\bf k}+ {\bf q}} )}{i \Omega -\left(\xi_{ {\bf k} + {\bf q} }-\xi_{ {\bf k} }\right)}\Bigg]\;,
\label{eq:local_bubble}
\end{align}
with $f(x)=1/(\exp[x/T]+1)$ the Fermi function and $\xi_{\mathbf{k}}$ the fermion lattice dispersion. 
For a 3d cubic lattice, for instance
\begin{align}
\xi_{\mathbf{k}}=-2 t\left(\cos k_x + \cos k_y + \cos k_z\right) - \mu\;,
\end{align}
with $t$ the nearest-neighbor hopping matrix element and $\mu$ the chemical potential which fixes the lattice filling ($\mu=0$ for half-filling, 
that is, $n=1/2$). 

From the free energy functional (\ref{eq:free_energy}) we will extract the phase boundary between the Fermi liquid and the metallic quantum charge glass and the associated quantum-critical dynamics of the density correlations. 
Minimization of Eq.~(\ref{eq:free_energy}) with respect to $q_{\rm EA}$, $\lambda$ and $D(\Omega)$ for each $\Omega$, yields a 
set of coupled saddle-point equations. The derivative with respect to $D(\Omega)$ for $\Omega>0$ together with the 
derivative with respect to $q_{\rm EA}$ requires the density response to obey the selfconsistency relation
\begin{align}
D(\Omega)=\int\frac{ d^d \mathbf{q}}{\left(2\pi\right)^d}
\frac{1}
{\left[-\Pi^{(0)}(\Omega,\mathbf{q})\right]^{-1}+ 2\lambda-J^2(\Omega)D(\Omega)}\;.
\label{eq:quadratic}
\end{align}
The lefthand side, when written as a geometric series, can be seen to express the self-consistent resummation of all cactus diagrams in the interactions $V_{ij}$. 
Minimization of Eq.~(\ref{eq:free_energy}) with respect to $\lambda$ gives back the global constraint on density fluctuations:
\begin{align}
T\sum_{\Omega}D(\Omega)+  q_{\text{EA}} 
=n(1-n)  \;.
\label{eq:chi_constraint}
\end{align}
Finally, the minimization with respect to $D(0)$ determines the Edwards-Anderson parameter self-consistently: 
\begin{align}
q_{\text{EA}}=
\int\frac{ d^d \mathbf{q}}{\left(2\pi\right)^d} 
\frac{J^2(0) q_{\text{EA}}}
{\left(\left[-\Pi^{(0)}(0,\mathbf{q})\right]^{-1}+ 2\lambda-J^2(0)D(0)\right)^2}\;.
\label{eq:q_saddle}
\end{align}
For vanishing interactions, $J(\Omega)=J(0)=0$, we have $q_{\text{EA}}=0$ and Eqs.~(\ref{eq:quadratic},\ref{eq:chi_constraint})  have the free fermion solution:
\begin{align}
D^{(0)}(\Omega)&= \int\frac{ d^d \mathbf{q}}{\left(2\pi\right)^d}\left[-\Pi^{(0)}(\Omega,\mathbf{q})\right]\nonumber\\
\lambda&=0\,.
\label{eq:bare}
\end{align}
Indeed, for free fermions, the constraint $T\sum_{\Omega} D^{(0)} (\Omega)= n \left(1-n\right)  $ is automatically fulfilled, 
and thus $\lambda(J=0)=0$.

Note that Eq.~(\ref{eq:q_saddle}) is a variant of the general glass instability condition:
\begin{eqnarray}
 \frac{J^2}{N}\sum_{ij}  \hat{\chi}^2_{ij}(\Omega=0) = 1
\label{eq:glassinstability}
\end{eqnarray}
where $\hat{\chi}$ is the full density-density correlator. In Appendix \ref{app:MillerHuse}, we present 
an alternative route to derive Eqs.~(\ref{eq:quadratic}-\ref{eq:q_saddle},\ref{eq:glassinstability}).

\subsection{Numerical results for metallic glass: phase diagram and density response}
\label{subsec:glasstransition}

\begin{figure}
\vspace*{2mm}
\includegraphics*[width=86mm,angle=0]{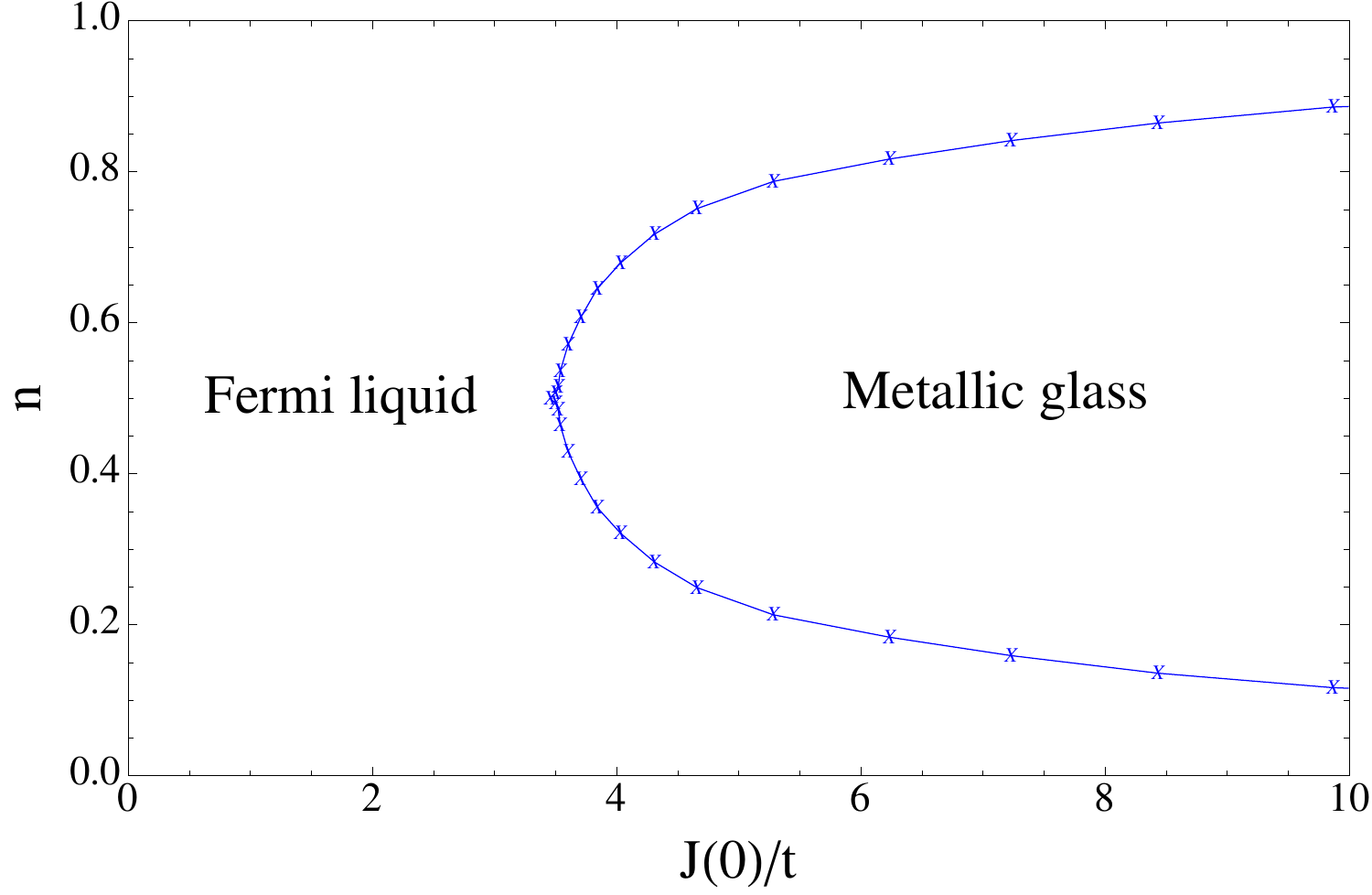}
\vspace*{-4mm}
\caption{(Color online) Numerically computed phase boundary of the metallic glass for various 
densities and ${\tilde W}=0$. The crosses correspond to data points computed from a simultaneous solution of 
the saddle-point equations~(\ref{eq:quadratic},\ref{eq:chi_constraint},\ref{eq:instability}); they are connected 
as a guide to the eye. Note that the phase boundary to the insulating glass is not plotted (see Fig.~\ref{fig:density_phasediag} for 
an illustration of both phase boundaries).}
\label{fig:glasstransition}
\end{figure}

\begin{figure}
\vspace*{2mm}
\includegraphics*[width=88mm,angle=0]{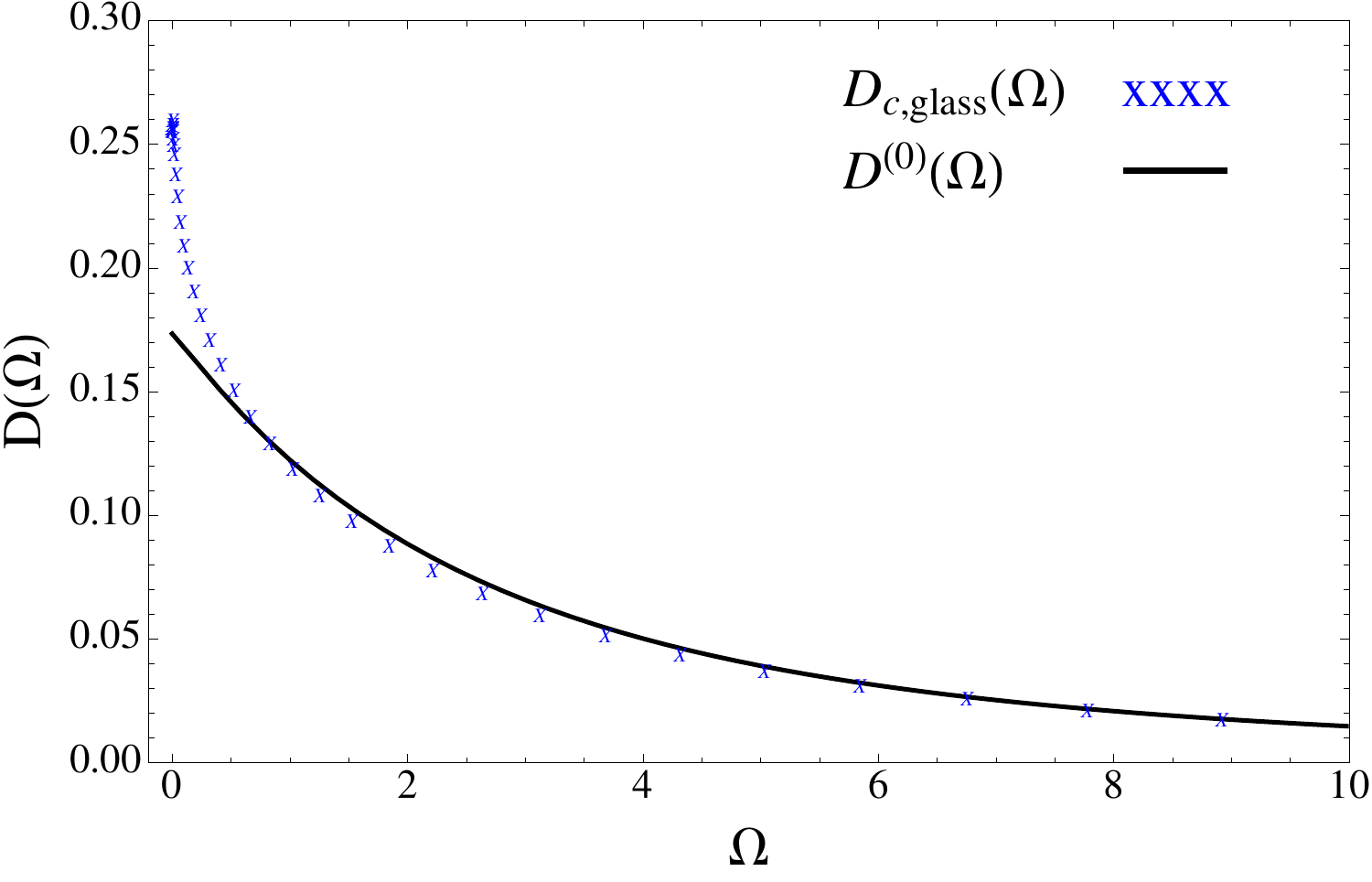}
\vspace*{-4mm}
\caption{(Color online) Comparison of the collective density response as a function of imaginary frequencies 
at the glass transition (blue crosses on the frequency grid 
computed from saddle-point equations~(\ref{eq:quadratic},\ref{eq:chi_constraint},\ref{eq:instability}), and its bare pendant 
from Eq.~(\ref{eq:bare}). The slope of the low energy response in the glass phase is singular, with a 
$\sqrt{|\Omega|}$ behavior for small frequencies.}
\label{fig:d_grid}
\end{figure}

We now compute the glass transition line (Fig.~\ref{fig:glasstransition}) and the associated density response (Fig.~\ref{fig:d_grid}), assuming no effective onsite disorder $\tilde W=0$ on a 3d-cubic lattice. We simultaneously solve Eqs.~(\ref{eq:quadratic}, \ref{eq:chi_constraint}) together with the criticality condition derived from Eq.~(\ref{eq:q_saddle}) 
(where it permits a solution with infinitesimal $q_{\text{EA}}$):
\begin{align}
1=
\int\frac{ d^d \mathbf{q}}{\left(2\pi\right)^d} 
\frac{J^2(0)}
{\left(\left[-\Pi^{(0)}(0,\mathbf{q})\right]^{-1}+ 2\lambda_{c,\text{glass}}-J^2(0)D(0)\right)^2}\;.
\label{eq:instability}
\end{align}
This criticality has an important consequence for the dynamic response. By writing $D(\Omega)=D(0)-\delta D_\Omega$, and expanding Eq.~(\ref{eq:quadratic}) around $\Omega=0$, we find that the condition (\ref{eq:instability}) entails a more singular low frequency behavior $\delta D_\Omega \sim \sqrt{|\Omega|}$, as compared to the behavior of the bare density response $\delta D^{(0)}_\Omega \sim |\Omega|$. This is illustrated by the explicit solution of $D(\Omega)$ on a frequency grid in Fig.~\ref{fig:d_grid}. 
Eq.~(\ref{eq:chi_constraint}) with finite $q_{\text{EA}}$, and more general arguments presented below ensure that this criticality extends into the glass phase. We will present an approximate analytical calculation of $D(\Omega)$ at the glass transition and in the metallic glass phase in the following subsection \ref{subsec:d_analytic}.


For the numerical solution of these equations we discretized $D(\Omega)$ on a frequency grid with varying step size up to a hundred grid points exploiting $D(\Omega)=D(-\Omega)$. 
We employed a modified version of Powell's Hybrid method algorithm for multi-dimensional root-solving \cite{gsl}. For the 3d-numerical momentum integrations of the right-hand-sides, we employed the VEGAS Monte Carlo algorithm \cite{gsl}. 
To avoid 6-dimensional integrations at each step of the multi-dimensional root solver, the momentum-integrated particle-hole 
bubble  was catalogued as 4-dimensional array and then 
``quadru-linearly'' interpolated in the integrands for the 3-dimensional $\mathbf{q}$-integration (see App.~\ref{app:phbubble} for 
some excerpts of $\Pi^{(0)}(\Omega,q_x,q_y,q_z)$).

The resulting phase boundary between Fermi liquid and metallic glass is plotted in Fig.~\ref{fig:glasstransition} for various fermion densities. 
Along the phase boundary the glass transition is of continuous nature, i.e. with a continuous onset of the Edwards-Anderson parameter $q_{\text{EA}}$ (in the case $\tilde W=0$). Like in quantum spin glasses, but in contrast to structural glasses (supercooled liquids), the dynamic freezing and the thermodynamic glass transition are thus expected to coincide. Accordingly one expects the replica symmetry breaking pattern within the glass phase to be continuous (full replica symmetry breaking) which ensures criticality and the existence of gapless collective modes~\cite{AndreanovMueller}.

One can see from Eq.~(\ref{eq:instability}) that as a function of density, the glass instability of the Fermi liquid occurs when $J\sim (\Pi^{0}(n)n^{1/2})^{-1}$. Using that $\Pi^{0}(n)\sim n/E_F(n) \sim n^{1-2/d}/t$, we infer that in dimensions $d=3$ the glass instability of the Fermi liquid scales as
\begin{align}
(J/t)_{\rm met} \sim n^{-3/2+2/d}=  n^{-5/6}
\label{Jt_met}
\end{align}
at low densities. This scaling with the fermion density is confirmed by the numerical results of Fig.~\ref{fig:glasstransition}.   
However, as we argued in Sec.~\ref{subsubsec:low_dens}, at lower densities (beyond the yellow 
tricritical points in Fig.~\ref{fig:phasediag_1}), this instability is pre-empted in equilibrium by a first order transition which takes place at $(J/t)_{\rm 1st} \sim n^{-1/2}\ll (J/t)_{\rm met}$. 
Nevertheless, the Fermi liquid phase should remain experimentally relevant even at these low densities, since it is very difficult to nucleate the localized insulator out of the metastable Fermi liquid phase, owing to the high nucleation barrier imposed by the long range interactions.

The glass transition, Fig.~\ref{fig:glasstransition}, and the associated emergence of a singular density response, Fig.~\ref{fig:d_grid}, 
are that of the universality class of infinite-range, metallic quantum Ising spin glasses. Those are characterized 
by the interaction- and disorder-induced freezing of Ising degrees of freedom in the presence of a metallic charge sector with gapless 
fermionic excitations, which damp the order parameter fluctuations. This was originally put forward in the context of metallic 
spin glasses by Sachdev, Read, and Oppermann \cite{sachdev95}, as well as by Sengupta and Georges \cite{sengupta95}. Later on this universality class was further analyzed in the form of a Landau theory for a mean field version of the electron-glass transition out of the Fermi liquid to by Dalidovich and Dobrosavljevi\'c \cite{dalidovich02}.

\subsection{Dynamic density response in the metallic glass}
\label{subsec:d_analytic}

It is instructive to derive an approximate analytical solution for $D(\Omega)$ neglecting 
the $\mathbf{q}$-dependence of the particle-hole bubble in Eq.~(\ref{eq:quadratic}). To this end we take $\Pi^{(0)}(\Omega,\mathbf{q})\rightarrow \Pi^{(0)}(\Omega,\mathbf{Q})$ with a fixed, prototypical
$\mathbf{Q}\neq 0 $ different from any potential nesting vectors, $|\mathbf{Q}|< 2 |\mathbf{k}_{\text{F}}|$. 
Then, Eq.~(\ref{eq:quadratic}) becomes a quadratic equation for the (approximate) density response, which we call $D_{\bf Q}(\Omega)$. It can be solved in closed form:
\begin{align}
D_{\mathbf{Q}}(\Omega)&=\frac{\left[-\Pi^{(0)}(\Omega,\mathbf{Q})\right]^{-1}+2 \lambda}{2 J^2(\Omega)}
-\nonumber\\&
\sqrt{\left(\frac{\left[-\Pi^{(0)}(\Omega,\mathbf{Q})\right]^{-1}+2 \lambda}{2 J^2(\Omega)}\right)^2-\frac{1}{J^2(\Omega)}}\;.
\label{eq:closed_quadratic}
\end{align}
where we chose the physical solution with minus-sign in front of the square-root, ensuring that $D_{\mathbf{Q}}(\Omega)$ 
decays to zero for large frequencies. Recall that the particle-hole bubble becomes zero for 
large external frequencies:
$\lim_{|\Omega|\rightarrow \infty} \Pi^{(0)}(\Omega,\mathbf{Q})\rightarrow 0$.
The criticality condition, Eq.~(\ref{eq:instability}) for fixed momentum $\mathbf{Q}$, applied to Eq.~(\ref{eq:closed_quadratic}), yields:
$\lambda_{c,\text{glass}}=J(0)-\frac{1}{2}\left[-\Pi^{(0)}(0,\mathbf{Q})\right]^{-1}$, 
and $D_{\mathbf{Q}}(0)_{c,\text{glass}}=\frac{1}{J(0)}$.
At the glass transition $D_{\mathbf{Q}}(\Omega)$ develops a singular response to small frequency pertubations of the local 
fermion density. 
The approximate energy scale 
\begin{align}
\Delta=\sqrt{\left(\lambda+\frac{\left[-\Pi^{(0)}(0,\mathbf{Q})\right]^{-1}}{2}\right)^2-J^2(0)}
\end{align}
controls the distance to the transition and separates two qualitatively distinct regimes in the dynamic density response of the Fermi liquid. 

Continuing Eq.~(\ref{eq:closed_quadratic}) to real frequencies, $i\Omega\rightarrow \Omega + i 0_{+}$, we obtain:
\begin{align}
-\text{Im}\,D_{\mathbf{Q}}(\Omega)\sim \left\{ 
\begin{array}{cl} \sqrt{\Omega}\;,\;\;\; & \Omega \gg \Delta\;,
\\ \Omega\;, & \Omega\ll \Delta\;\;.  \end{array}\right.
\label{eq:glass_compressi}
\end{align}
The self-consistency condition (\ref{eq:chi_constraint})  pins  $\Delta=0$
in the entire glass phase, and the density response remains singular at zero frequency. This property holds true independently of the above approximations, as  
can be formally derived assuming full replica symmetry breaking in the glass phase~\cite{AndreanovMueller}. 

The key ingredient for the unusual square-root behavior for both, $D_{\mathbf{Q}}(\Omega)$ and the full $D(\Omega)$ of 
Fig.~\ref{fig:d_grid}, is the coupling of the collective 
charge fluctuations to the metallic particle-hole continuum which {\em coexists} with glassy, amorphous density order. 
The low frequency behavior of the particle-hole bubble (see Fig.~\ref{fig:mod_omega} in the appendix) at finite momentum
transfer $\mathbf{Q}$:
\begin{align}
\Pi^{(0)}(\Omega,\mathbf{Q})-\Pi^{(0)}(0,\mathbf{Q})\sim |\Omega|\;,
\label{eq:mod_omega}
\end{align}
underlies Eq.~(\ref{eq:glass_compressi}) and also enters into the numerical computation of Fig.~\ref{fig:d_grid}. 

%
%

\subsection{Fermionic quasiparticles in the metallic glass} 

The single-particle properties of the underlying fermions in systems belonging to the universality class of
infinite-range, metallic quantum Ising glasses have been worked out by Sengupta and Georges \cite{sengupta95}.
At $T=0$, at the critical point and in the glass phase, the fermions remain well-defined quasi-particles. 
Indeed, the leading self-energy correction due to low frequency charge fluctuations scale as
\begin{align}
\Sigma_{f}(\omega)\sim\omega^{3/2}\;,
\end{align}
with a frequency exponent $>1$.
Nevertheless, at finite temperatures this translates into non-analytic corrections in the transport properties. 
The finite-temperature 
resistivity, for example, scales as $\delta \rho(T)\sim T^{3/2}$ in the quantum-critical regime above the QCP \cite{sengupta95} and in
the entire metallic quantum glass phase. Deeper in the glass phase, the metallic diffusion eventually breaks down when the localization transition to the Anderson-Efros-Shklovskii glass of 
Sec.~\ref{sec:localized} is reached.

   
\section{Phase boundaries of the metallic glass at $\widetilde{W}> J$}
\label{sec:fractal}

In the preceding section we have argued for a metallic glass phase at moderate densities and weak external disorder.
However, also  when the effective onsite disorder $\widetilde{W}$ is larger than the infinite-range interaction $J$, we expect an intermediate metallic glass phase ``strip'' between the Fermi liquid and 
the AES glass, as shown in Fig.~\ref{fig:phasediag_1}. Below we will present scaling arguments to justify this scenario, with both transitions at the borders of the metallic glass ``strip'' (the metal-insulator transition on ``top'' and the glass transition 
on the ``bottom'') being continuous.

For vanishing interactions $J=0$, the two instability lines must join at the Anderson transition of the disordered free fermions, where 
$\widetilde{W}_{\text{loc}}/t = W_{\text{loc}}/t =21.29$ (for Gaussian disorder at half filling~\cite{Ohtsuki}). In the presence of weak interactions, the frozen fields from the glassy density order tend to increase the disorder variance by $\delta (W^2) \lesssim  J^2$, which has a similar effect as a weak increase of disorder $\delta W \sim  J^2/W$. Accordingly one expects the critical value $W/t$ for delocalization to decrease by a relative amount 
\begin{eqnarray}
\delta \left[\frac{W}{t}\right]_{\rm loc} \equiv \left[\frac{W}{t}\right]^{(0)} - \left[\frac{W}{t}\right]_{\rm loc}^{(J)}  \,\sim -\left(\frac{J}{W}\right)^2,
\label{shiftloc}
\end{eqnarray}
since stronger hopping is necessary to compensate for the extra disorder.

\subsection{The role of fractality for the glass transition}
On the other hand, we have to analyze the glass instability (\ref{eq:glassinstability}) within the metallic phase. In the limit $J\to 0$, it holds that  $\hat \chi\to \chi(J=0)$, which reduces to the (exact) susceptibility of non-interacting fermions. As noted in Ref.~\onlinecite{dobro03}, the sum over susceptibility squares,
\begin{eqnarray}
\chi_2 \equiv \frac{1}{N} \sum_{ij}  \overline{\chi^2_{ij}(\Omega=0)}
\label{chi2}
\end{eqnarray}
diverges at the Anderson transition. This implies a glass instability already within the metallic phase, even for very weak interactions $J\ll t,W$. 
To the best of our knowledge, the precise divergence of $\chi_2$, or equivalently, of $\overline{\chi_{ij}^2}$, as $\left[\frac{W}{t}\right] \to \left[\frac{W}{t}\right]_{\rm loc}^{(0)}$, is not known. However, the disorder-averaged density-density correlator has been well studied, since it reveals interesting properties of the fractal nature of the electronic wavefunctions and the anomalous diffusion at the Anderson transition~\cite{RMPMirlin}. 
In particular, the spatial Fourier transform of $\overline{\chi_{ij}(\Omega=0)}$ behaves as 
\begin{eqnarray}
\overline{\chi_{q}(\Omega\to 0)} \equiv D(q,\Omega\to 0)\sim \frac{1}{\delta_\xi f(|q|\xi)}\;,
\end{eqnarray}
where $\xi$ is the correlation length, which diverges at the Anderson transition as $\xi\sim (1 -t/t_c)^{-\nu}$; $\delta_\xi= 1/\nu\xi^d$ is the single particle level spacing in the correlation volume and $\nu$ is the density of states. The scaling function $f(x)$ behaves as~\cite{chalker88, kravtsovYevtushenko}
\begin{eqnarray}
f(x)\sim  x^{d_2}, \quad x\gg 1,\\
f(x)\sim  x^{2}, \quad x\ll 1.
\end{eqnarray}
Here, $d_2$ is the fractal dimension associated with the distribution of $|\psi^4(x)|$ over all space, $\psi(x)$ being a critical single particle wavefunction at the delocalization transition.
From this one obtains that
\begin{eqnarray}
\hat \chi_2 &\equiv &\frac{1}{N} \sum_{ij}  \overline{\chi_{ij}(\Omega=0)}^2 = \int \frac{d^dq}{(2\pi)^d} \overline{\chi_{q}(\Omega= 0)}^2 \nonumber\\
&\sim &  \xi^{2(d-d_2)} \sim (1 -t/t_c)^{-\hat\alpha}\;.
\end{eqnarray}
The exponent results as $\hat \alpha= 2\nu(d-d_2)\approx 4$, using the known values $d_2[d=3] \approx 1.3$ ~\cite{Mildenberger, RMPMirlin} and  $\nu= 1.57$.~\cite{Ohtsuki}

It is reasonable to expect that $\chi_2\sim (1-t/t_c)^{-\alpha}$ needed in Eq.~(\ref{chi2}) diverges at least as fast as $\hat\chi_2$, and thus $\alpha\geq \hat \alpha$. Hence we expect that the glass instability is displaced from the non-interacting Anderson transition by an amount
\begin{eqnarray}
\delta \left[\frac{W}{t}\right]_{\rm met} \equiv \left[\frac{W}{t}\right]^{(0)} - \left[\frac{W}{t}\right]_{\rm met}^{(J)}\, \sim -\left(\frac{J}{W}\right)^{2/\alpha}.
\end{eqnarray}
Comparing with Eq.~(\ref{shiftloc}) and noting that $\alpha>1$, one sees that as $J\to 0$ the glass instability line approaches the non-interacting Anderson transition from smaller values of $(W/t)$ than the delocalization instability, as indicated in Fig.~\ref{fig:phasediag_1}. These arguments confirm the existence of an intermediate phase between disordered Fermi liquid and Anderson localized glass, which exhibits both metallic diffusion and glassy density order. 


\section{Conclusion}
\label{sec:conclu}

Based on the calculations and arguments of this paper and a previous work \cite{strack11}, we believe that many-body cavity QED could 
evolve into a new platform to explore the physics of long-range quantum glasses. The long range of the photon-mediated interactions 
simplifies the theoretical analysis for these systems; this may allow for quantitative comparison between experiment and theory. 
Our work complements previous proposals on glassy many-body systems in quantum optics
 \cite{damski03,ahufinger05} which have focussed onto the creation of onsite disorder or random short range exchanges by using random optical lattices and species admixture. 
  
In Ref.~\onlinecite{strack11}, we predicted a quantum spin glass transition of fixed, stationary atoms 
where the source of quantum fluctuations was 
spontaneous tunneling between suitable chosen internal states of the atoms. In the present paper, we considered itinerant fermions where the source of quantum fluctuations is tunneling between adjacent lattice sites. We found a metallic glass phase with a gapless 
Fermi sea in the presence of random density order and, for stronger interactions, a localized state with strongly random charge distributions 
and vanishing conductivity at $T=0$. 

Various aspects of the phase transitions into and out of these glass phases are worth further studies: 
One is to quantify further the impact of the fractality of wavefunctions close to the Anderson transition and compute 
the critical exponent $\alpha$ with which the glass instability line approaches the non-interacting Anderson transition point.
%
%
Experimentally, repeated ramps from the Fermi liquid phase into the glass phase should produce 
metastable states with a different density pattern at each run. The resulting variations in absorption images 
could be particularly strong, and therefore perhaps easier to detect, 
in the low density regime where the glass and localization transition is of first order.

What happens when one considers models of type Eq.~(\ref{eq:fermi_hamiltonian_1}) with bosonic atoms? 
Numerical simulations \cite{carleo09,tam10} and 
replica calculations \cite{yu11} for similar models agree on the existence of a stable ``superglass'' phase with superfluid phase coherence in the presence of glassy random density order. It is tempting to identify the superglass as the bosonic pendant to the metallic fermion glass. 
However, at finite temperatures, the domain of the superglass (cf. Fig. 1 of Ref.~\onlinecite{yu11}) is reduced because the main 
effect of thermal fluctuations is to weaken the superfluid phase coherence of the bosons. 
For fermions, on the other hand, the domain of the 
metallic glass increases at finite $T$ because the main effect of thermal fluctuations is to enhance the inelastic scattering rate and to weaken the localizing disorder potential. 

We described the disordered atom-light quantum phases using
an effective equilibrium ground-state description. In actual optical cavities, one deals with pumped, steady-state phases certain features 
of which may not be captured in an effective equilibrium description \cite{dimer07,nagy11,bhaseen12}. We hope to address the non-equilibrium properties of open, disordered Dicke models in the near future in a forthcoming paper.
We also want to analyze how the intriguing properties of classical and quantum glasses such as aging 
(see Ref.~\onlinecite{kennett01} and references therein), or out of equilibrium dynamics and avalanches~\cite{Pazmandi}
are modified in the open, driven, steady-states in many-body cavity QED.

Finally, it will be interesting to investigate how the results of this paper and Ref.~\onlinecite{strack11} scale with the number of cavity modes and the choice of cavity mode profiles.

\begin{acknowledgments}

We thank Alexei Andreanov, Vladimir Kravtsov, Giovanni Modugno, Francesco Piazza and Giacomo Roati for useful discussions and Manuel Schmidt 
for bringing Ref.~\onlinecite{kosterlitz76} to our attention. This research was supported by the DFG under grant Str 1176/1-1, by the NSF under Grant DMR-1103860, and by a MURI grant from AFOSR.

\end{acknowledgments}

\appendix

\section{Alternative derivation of the glass instability equations (\ref{eq:quadratic},\ref{eq:instability},\ref{eq:glassinstability})} 
\label{app:MillerHuse}

\subsection{Replica approach}
Here we formally derive the glass instability for arbitrary onsite disorder using the replica approach. We start from model (\ref{eq:fermi_hamiltonian}). Replicating $n$ times and taking the disorder average and introducing Hubbard Stratonovich fields to decouple the quartic density interactions, one obtains the replicated partition function 
\begin{align}
\overline{Z^n}= \int \prod_{a,i,\tau}{\mathcal D}\bar{c}_i^a(\tau){\mathcal D}c_i^a(\tau) \prod_{a\leq b,\tau,\tau'}{\mathcal D}{\mathcal Q}_{ab}(\tau,\tau')\exp[-{\mathcal S}_{\mathcal {\mathcal Q}}-{\mathcal S}_c]
\end{align}
with the action

\begin{align}
\mathcal{S}_c&=\sum_{a=1}^n\int_0^{1/T} d\tau \sum_{i=1}^N
\bar{c}_i^a (\tau) \left(\partial_{\tau}  -\mu\right) c^a_{i}(\tau)
\nonumber\\
&+\sum_{a=1}^n \int_0^{1/T} d\tau \sum_{\langle i, j\rangle}
t\left( \bar{c}^a_i (\tau) c^a_j(\tau) + \bar{c}^a_j (\tau) c^a_i(\tau)\right) \nonumber\\
&-\frac{1}{2}\sum_{a,b=1}^n \int_0^{1/T} d\tau\, d\tau' \left[ W^2+J^2(\tau-\tau') {\mathcal Q}_{ab}(\tau,\tau')\right]n_i^a(\tau)n_i^b(\tau')\;,\nonumber\\
\mathcal{S}_{{\mathcal Q}}&=\frac{N}{4}\sum_{a,b=1}^n \int_0^{1/T} d\tau \, d\tau'  J^2(\tau-\tau') {\mathcal Q}_{ab}^2(\tau,\tau')
\label{replicaaction}\;.
\end{align}

In the thermodynamic limit ($N\rightarrow \infty$), the infinite range of the interactions allows us to take the saddle point with respect to ${\mathcal Q}_{ab}$, which satisfy the saddle point equations
\begin{align}
{\mathcal Q}_{ab}(\tau,\tau') = \frac{1}{N} \sum_i \langle n_i^a(\tau)n_i^b(\tau') \rangle\;.
\end{align}
(Note the slightly different definition of ${\mathcal Q}$ as compared to $Q_{ab}$ defined in Eq.~(\ref{Qab}).)
From here on we neglect the retardation in the mediated coupling $J(\tau-\tau')$ and replace it simply with $J$.  
As usual, the saddle point values of ${\mathcal Q}_{ab}$ are independent of $\tau,\tau'$ for $a\neq b$, and depend only on $\tau-\tau'$ for the replica diagonal ${\mathcal Q}_{aa}$.
In the disordered, replica symmetric phase the term $\tilde W^2 \equiv W^2+J^2 {\mathcal Q}_{a\neq b}$ can be recognized as the variance of the self-consistent effective disorder (\ref{effdis}). Note that since ${\mathcal Q}_{a\neq b}\geq \langle n\rangle ^2$, the effective disorder never really vanishes, unless $\epsilon_i$ and $V_{ij}$ have special correlations. 

To detect a glass instability one has to solve first the saddle point equations of the disordered phase. This yields a replica symmetric solution ${\mathcal Q}^{\rm RS}_{ab}(\tau,\tau')$ which encodes the Edwards Anderson overlap
\begin{align}
q_{\rm EA}\equiv {\mathcal Q}_{a\neq b}^{\rm RS} = \frac{1}{N} \sum_i \langle n_i \rangle^2\;,
\end{align}
and the average local susceptibility,
\begin{align}
{\mathcal Q}_{aa}^{\rm RS}(\tau-\tau') -q_{\rm EA} = \frac{1}{N} \sum_i \langle n_i(\tau)n_i(\tau') \rangle_c\equiv \chi_{\rm loc}(\tau-\tau')\;.
\end{align}
 
The glass instability is found by writing ${\mathcal Q}={\mathcal Q}^{\rm RS}+\delta Q$ and expanding the free energy in $\delta Q$. A standard cumulant expansion yields 
\begin{align}
\overline{Z^n}[{\mathcal Q}^{\rm RS}+\delta Q] = \exp[- N(\beta F^{\rm RS}+ \delta(\beta F))]\;,
\end{align}
 which by virtue of the extremality of ${\mathcal Q}^{\rm RS}$ starts with a quadratic term.
The glass instability is signalled by the vanishing of the coefficient of the term $ \delta Q_{ab}^2$ (the mass of the replicon mode),
\begin{align}
0&=(\beta J)^2-\frac{J^4}{N}\sum_{i,j}  \left\langle \int d\tau d\tau'  n_{i,a}(\tau)n_{j,a}(\tau')\right\rangle_c^2 \nonumber\\
\quad & \equiv (\beta J)^2\left(1-\frac{J^2}{N}\sum_{i,j}\hat\chi_{ij}^2\right)\;. 
\label{eq:repliconmass}
\end{align}
Hereby the correlator 
\begin{align}
\hat\chi_{ij} = \int d\tau  \left\langle n_{i,a}(\tau)n_{j,a}(0)\right\rangle_c \;, 
\end{align}
has to be evaluated with the replica symmetric action. Eq.~(\ref{eq:repliconmass}) 
is to be compared with Eqs.~(\ref{eq:glassinstability},\ref{eq:instability}).

\subsection{Cavity approach: local selfconsistent action}
In this subsection we derive the glass instability 
conditions from a selfconsistent local action derived within a cavity approach. We start from the action (\ref{eq:fermi_bose_action}), and split it into single particle and interaction part $\mathcal{S} = \mathcal{S}_1+ \mathcal{S}_{V}$,
\begin{align}
\nonumber\\
\mathcal{S}_1&=\int_0^{1/T} d\tau \sum_{i=1}^N
\bar{c}_i (\tau) \left(\partial_{\tau}+ \tilde \varepsilon_i -\mu\right) c_{i}(\tau)
\nonumber\\
&-\int_0^{1/T} d\tau \sum_{\langle i, j\rangle}
t\left( \bar{c}_i (\tau) c_j(\tau) + \bar{c}_j (\tau) c_i(\tau)\right)\;, \nonumber\\
\mathcal{S}_{V}&=-\frac{1}{2}\sum_{i,j=1}^N \int_0^{1/T} d\tau \int_0^{1/T} d\tau'
V_{ij}(\tau-\tau')\times\nonumber\\
&\quad\quad \quad\quad \quad\quad 
(n_i(\tau)-\langle n_i\rangle)( n_j(\tau')-\langle n_j\rangle)
\label{eq:fermi_bose_action2}\;,
\end{align}
where $n_i(\tau) \equiv \bar{c}_i (\tau){c}_i (\tau)$, and 
\begin{align}
\tilde \varepsilon_i = \varepsilon_i-\sum_{j\neq i} V_{ij} (\Omega=0)\, \langle n_j\rangle.
\label{extradis}
\end{align}

We now take advantage of the infinite range nature of the interactions, to transform the above problem {\em exactly} into a {selfconsistent} single-site 
problem with retarded density-density interactions.
The extra contribution to the disorder (\ref{extradis}) can be treated as an additional Gaussian disorder with variance $J^2 \overline{\langle n_i \rangle^2}= J^2 \left(n^2 +\overline{\langle n_i \rangle^2}^c\right)$. This type of disorder is essentially unavoidable in the system. Note that  if it does not vanish, $\tilde \varepsilon_i\neq 0$, there are density inhomogeneities already in the disordered phase, i.e., $\langle n_i\rangle \neq \langle n_j\rangle$ for $i\neq j$. Unfortunately this renders the exact evaluation of the self-consistency problem very hard, and one has to resort to approximations in order to obtain quantitative results. 

Upon average over $V_{ij}$ a subsequent Hubbard Stratonovich transformation and a saddle point approximation, one finds that the interaction term can be resummed as
\begin{align}
\mathcal{S}_{V}'= \sum_i (n_i(\tau)-\langle n_i\rangle) R(\tau-\tau') ( n_j(\tau')-\langle n_j\rangle), 
\end{align}
with the kernel 
\begin{align}
R(\Omega) &= 
J^2(\Omega)\overline{ \frac{1}{N}\sum_j  \langle n_j(\Omega)n_j(-\Omega) \rangle_{\mathcal{S}_{1}+\mathcal{S}_{V}'}}  \equiv 
J^2(\Omega) \chi_{\rm loc}(\Omega),
\end{align}
where the average local susceptibility $\chi_{\rm loc}$ must be found self-consistently. 

When computing the density-density correlator 
\begin{align}
\hat\chi_{ij}(\Omega) \equiv \langle n_i(\Omega)n_j(-\Omega) \rangle_{\mathcal{S}_{1}+\mathcal{S}_{V}'},  
\end{align}
with the self-consistent action, one should be aware, however, that the above resummation does not include terms which contain a given coupling $V_{ij}$ only once within the expansion in interactions. Those are indeed irrelevant upon site or disorder averaging. However, they cannot be neglected when the glass susceptibility is computed,
\begin{align}
\chi_{\rm glass} 
 = \frac{1}{N}\sum_{ij} \chi_{ij}^2(\Omega=0).
\end{align}
A glass transition manifests itself by a divergence of this susceptibility, which in mean field systems signals the emergence of ergodicity breaking in the form of many pure states, and replica symmetry breaking.
In this formula $\chi_{ij}$ is the full density-density correlator. To the relevant order in $V_{ij}$ it can be obtained from $\hat \chi_{ij}$ by the summation of the geometric series, 
\begin{align}
\chi_{ij} = \hat \chi_{ij} + \sum_{k, \ell} \hat \chi_{ik} V_{k\ell}   \hat \chi_{\ell j}+... \;.
\end{align}
Upon taking the square and summing over all pairs of sites one finds
\begin{align}
\chi_{\rm glass} = \frac{1}{1- \frac{J^2}{N}\sum_{ij} \hat \chi_{ij}^2(\Omega=0)},
\label{criticality}
\end{align}
which diverges when the denominator vanishes, reproducing Eq.~(\ref{eq:repliconmass}).

In mean field glasses, the continuous breaking of replica symmetry ensures usually that the condition (\ref{criticality}) remains fulfilled, even within the glass phase, where $\hat \chi_{ij}$ is to be interpreted as the  density-density correlator within one metastable state (and contributions from single couplings dropped). This was explicitly shown in the case of the infinite range transverse field Ising spin glass~\cite{AndreanovMueller}. This phenomenon of maintained marginal stability is at the basis of the permanent gaplessness of the quantum glass phase.

\subsection{Generalized Miller-Huse type analysis}
Following a reasoning similar to the one by Miller and Huse~\cite{miller93} for the transverse field spin glass, we formally compute the local density-density correlator $\hat \chi$ in a perturbation series in the interaction action ${\mathcal S}_V$. The perturbation series for the correlator $\hat \chi$ can be formally summed up as a geometric series of local interactions $R(\tau-\tau')$ linking "proper polarizability" blobs $\Pi_{ij}(\tau-\tau')$ that cannot be reduced into two separate pieces by cutting a single  $R$-line, 
\begin{align}
\hat\chi_{ij}(\Omega)& = \Pi_{ij}(\Omega) + \sum_k \Pi_{ik}(\Omega) R(\Omega) \Pi_{kj}(\Omega)+... \nonumber\\
&= \left[ \frac{1}{{\mathbf \Pi}^{-1}(\Omega)-R(\Omega)}\right]_{ij}.
\label{geomseries}
\end{align}
The proper polarizability has itself a power series expansion in $R$. To lowest order one has simply 
\begin{align}
\Pi_{ij}(\Omega)=\chi^{(0)}_{ij}(\Omega)+O(R),
\label{lowestorderPi}
\end{align}
where $\chi^{(0)}$ denotes the non-interacting density-density correlator,
\begin{align}
\langle n_i(\Omega) n_j(-\Omega') \rangle_{J=0} =:\chi^{(0)}_{ij}(\Omega) \delta_{\Omega,\Omega'}.
\end{align}

As derived above in Eqs.~(\ref{eq:repliconmass},\ref{criticality}), the glass transition occurs when 
\begin{align}
1 = \frac{J(0)^2}{N} {\rm Tr} \left[{\mathbf \Pi}^{-1}(\Omega=0)-R(\Omega=0) \right]^{-2}.
\end{align}
In the approximation where we neglect the effective disorder, $\tilde\varepsilon_i=0$, in the quantum disordered phase, one has translational invariance, which allows one to work in Fourier space.

The average local density-density correlator,
\begin{align}
D(\Omega)=\frac{1}{N} \sum_i  \hat \chi_{ii}(\Omega)= \frac{1}{N} {\rm Tr} \left[{\mathbf \Pi}^{-1}(\Omega)-R(\Omega) \right],
\end{align}
must satisfy the selfconsistency condition
\begin{align}
R(\Omega)  =  J^2(\Omega) D(\Omega).
\end{align}
It must also obey the constraint
\begin{eqnarray}
&&\int \frac{d\Omega}{2\pi} D(\Omega) = \frac{1}{N}\sum_i \overline{(n_i(\tau) -\langle n_i \rangle)^2}
\label{constraintonD}
\\
&&\quad = \frac{1}{N}\sum_i \langle n_i \rangle (1-\langle n_i \rangle) = n(1-n) - [\overline {\langle n_i \rangle^2}-n^2].\nonumber
\end{eqnarray}
In the absence of onsite disorder, the last term in brackets vanishes. The constraint (\ref{constraintonD}) is fulfilled automatically by an exact solution. However, if $D$ is evaluated within an approximate scheme, e.g. with the help of Eqs.~(\ref{geomseries},\ref{lowestorderPi}), one should impose this constraint to obtain a better approximation.
In particular, we can satisfy the short time constraint (\ref{constraintonD}) by adding an adjustable short-time component $\lambda$ to the relation 
$J^2(\Omega)(D(\Omega)-\lambda)=R(\Omega)$, to correct for the errors at high frequencies introduced by the approximations involved in computing $D$. The better the approximation, the smaller will be the $\lambda$ required to enforce the short time constraint. This recipe turns out to be essentially equivalent to the global constraint we introduced in Sec.~\ref{SPfreeenrgy}.

We thus have to solve simultaneously 
\begin{align}
\int \frac{d\Omega}{2\pi} D(\Omega) = n(1-n) - [\overline {\langle n_i \rangle^2}-n^2],
\end{align}
and
\begin{align}
D(\Omega) = \frac{1}{N} {\rm Tr} \left( \frac{1}{{\mathbf \Pi}^{-1}(\Omega)-J^2(\Omega)[D(\Omega)-\lambda]} \right).
\label{SC}
\end{align}
The glass transition arises when 
\begin{align}
1 & = \frac{J(0)^2}{N} {\rm Tr} \left( \frac{1}{{\mathbf \Pi}^{-1}(0)-J^2(0)[D(0)-\lambda]} \right)^2 \nonumber\\
&= \frac{\partial}{\partial[D(0)]} \frac{1}{N} {\rm Tr} \left( \frac{1}{{\mathbf \Pi}^{-1}(0)-J^2(0)[D(0)-\lambda]} \right)\;.
\end{align}
The latter relation leads to a singular behavior of $D(\Omega)$ around $\Omega\to 0$, as one may see by expanding Eq.~(\ref{SC}) around $\Omega=0$, very similarly as in quantum spin glasses~\cite{miller93, AndreanovMueller} . This singularity ensures the presence of spectral weight ${\rm Im}D(\Omega\to \omega+i\delta)$ at all finite real frequencies $\omega$.    

The three last equations are to be compared with Eqs.~(\ref{eq:quadratic}-\ref{eq:q_saddle}) to which they reduce in the translationally invariant case. 

\section{Particle-hole bubble as function of external frequency and momenta}
\label{app:phbubble}

\begin{figure}
\vspace*{0mm}
\includegraphics*[width=82mm]{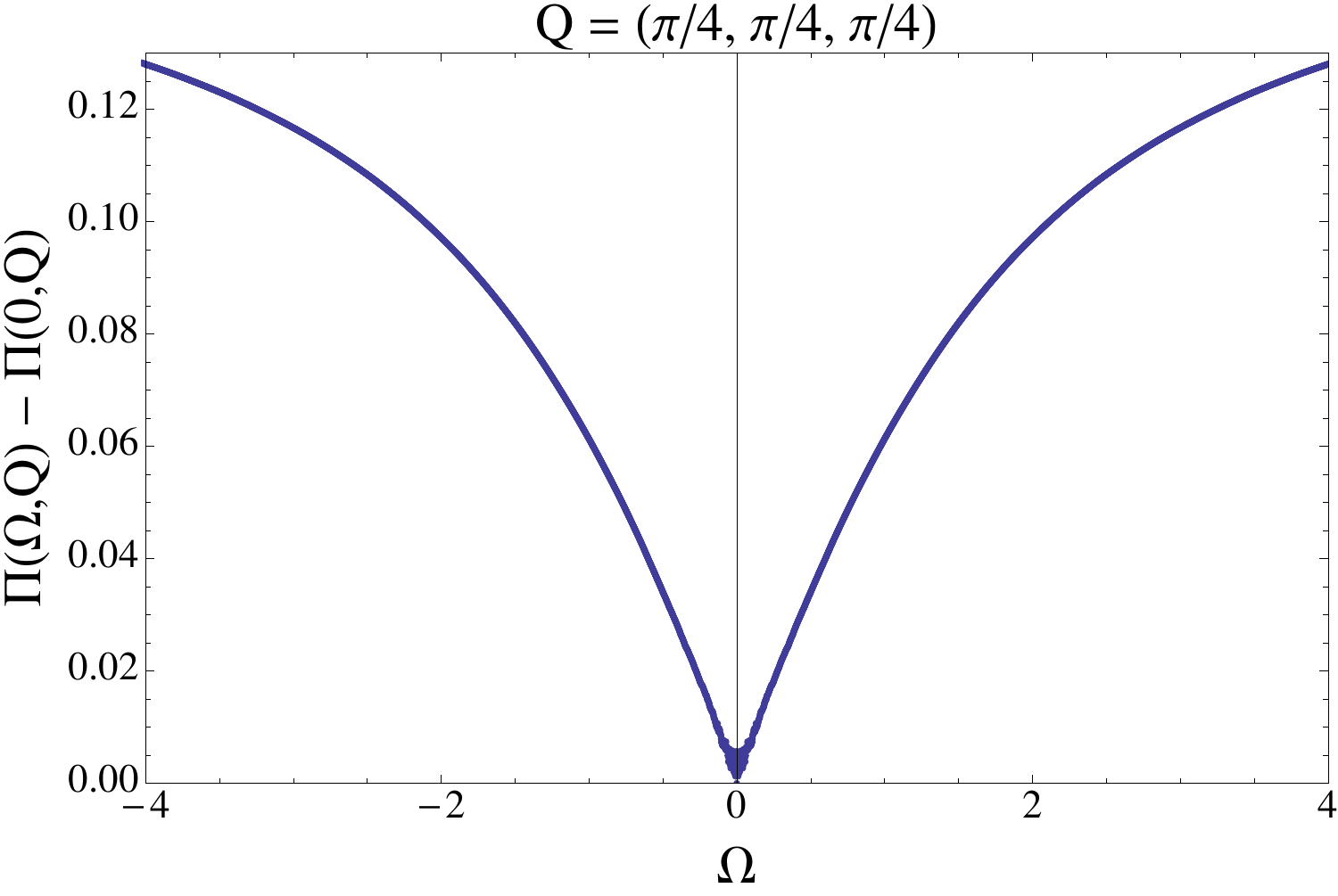}
\vspace*{7mm}
\caption{(Color online)  $|\Omega|$ behavior of the particle-hole bubble, Eq.~(\ref{eq:local_bubble}), as a function 
of external frequency for fixed momentum transfer $\mathbf{Q}$.}
\label{fig:mod_omega}
\end{figure}

In Fig.~\ref{fig:mod_omega}, we display the particle-hole bubble as a function 
of external frequency for fixed momentum transfer. The low-frequency part 
behaves as $~|\Omega|$ as alluded to in Eq.~(\ref{eq:mod_omega}).

\begin{figure}
\vspace*{2mm}
\includegraphics*[width=86mm,angle=0]{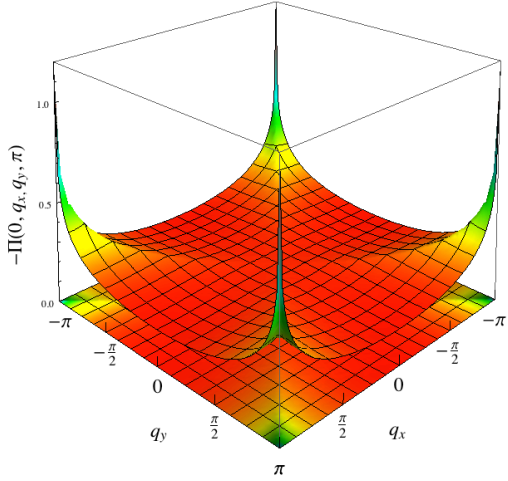}
\vspace*{-6mm}
\caption{(Color online) Exemplary momentum behavior of the static ($\Omega=0$) 
particle-hole bubble, Eq.~(\ref{eq:local_bubble}), at half-filling in 3 dimensions.}
\label{fig:ph_bubbles}
\end{figure}

In Fig.~\ref{fig:ph_bubbles}, we plot an exemplary $\Omega=0$ bubble as function of momenta as 
occurring in the saddle-point equations (\ref{eq:quadratic},\ref{eq:chi_constraint},\ref{eq:instability}).
As expected, the dominant contributions 
come from momenta in the vicinity of the nesting condition $\mathbf{q}\approx\mathbf{Q}_{\text{nest}}=(\pi,\pi,\pi)$.
Although logarithmically divergent at $\mathbf{q}=\mathbf{Q}_{\text{nest}}$, the right-hand-side of the saddle-point equations 
remains regular as it involves an additional 3-dimensional integration over $\mathbf{q}$.

\section{Landau expansion}
Let us start again from the action
\begin{align}
\nonumber\\
\mathcal{S}_1&=\int_0^{1/T} d\tau \sum_{i=1}^N
\bar{c}_i (\tau) \left(\partial_{\tau}+ \tilde \varepsilon_i -\mu\right) c_{i}(\tau)
\nonumber\\
&-\int_0^{1/T} d\tau \sum_{\langle i, j\rangle}
t\left( \bar{c}_i (\tau) c_j(\tau) + \bar{c}_j (\tau) c_i(\tau)\right)\;, \nonumber\\
\mathcal{S}_{V}&=-\frac{1}{2}\sum_{i,j=1}^N \int_0^{1/T} d\tau \int_0^{1/T} d\tau'
V_{ij}(\tau-\tau')\times\nonumber\\
&\quad\quad \quad\quad \quad\quad 
(n_i(\tau)-n)( n_j(\tau')-n)
\label{eq:fermi_bose_action2}\;,
\end{align}
where 
\begin{align}
\tilde \varepsilon_i = \varepsilon_i-\sum_{j\neq i} V_{ij} (\Omega=0)\, n;, 
\label{extradis}
\end{align}
Here, we subtract the average density, not the inhomogeneous local density.
Upon replicating and disorder averaging one gets the action (neglecting any frequency dependence of $V$)
\begin{align}
\nonumber\\
\mathcal{S}_1&=\int_0^{1/T} d\tau \sum_{a,i=1}^N
\bar{c}_{i,a} (\tau) \left(\partial_{\tau} -\mu\right) c_{i,a}(\tau)
\nonumber\\
&-\int_0^{1/T} d\tau \sum_{a,\langle i, j\rangle}
t\left( \bar{c}_{i,a} (\tau) c_{j,a}(\tau) + \bar{c}_{j,a} (\tau) c_{i,a}(\tau)\right) \nonumber\\
&-(W^2+J^2n^2)\int\int _0^{1/T} d\tau d\tau' \sum_{a,b,i} n_{i,a}(\tau)n_{i,b}(\tau')\nonumber\\
\mathcal{S}_{V}&=-\frac{J^2}{4?}\sum_{i,j=1}^N \int_0^{1/T} d\tau \int_0^{1/T} d\tau'
\nonumber\\
&\quad\quad \quad
(n_{i,a}(\tau)-n)( n_{i,b}(\tau')-n)(n_{j,a}(\tau)-n)( n_{j,b}(\tau')-n)\;,
\end{align}
Hubbard Stratonovich transforming the last term one gets the local interaction
\begin{align}
\mathcal{S}'_{V}&=-\frac{J^2}{4?}\sum_{i}^N \int_0^{1/T} d\tau \int_0^{1/T} d\tau'
\nonumber\\
&\quad\quad \quad
(n_{i,a}(\tau)-n)( n_{i,b}(\tau')-n)Q_{ab}(\tau,\tau')\;,
\end{align}
where $Q_{ab}(\tau,\tau')=\frac{1}{N} \sum_j \langle (n_{j,a}(\tau)-n)( n_{j,b}(\tau')-n) \rangle$. Unfortunately, for finite $\tilde W^2\equiv W^2 +n^2J^2$, $Q_{ab}$ is not small. However, if $\tilde W$ is small, we can expand in $Q_{ab}$. That is, we make a cumulant expansion in $Q_{ab}$. The prefactors of the terms $Tr Q^2$, $Tr Q^3$ and $\sum_{ab}Q_{ab}^3$ in the free energy functional are respectively
 \begin{align}
&C_{Tr Q^2}=\sum_{i,j} \langle \int d\tau d\tau'  (n_{i,a}(\tau)-n)( n_{j,a}(\tau')-n) \rangle^2 \equiv (\beta \chi_{ij})^2\nonumber\\
&C_{Tr Q^3}=\sum_{i,j,k}  (\beta \chi_{ij})(\beta \chi_{jk})(\beta \chi_{ki}) \\
&C_{Q_{ab}^3}=\sum_{i,j,k} \langle \int d\tau d\tau'd\tau''  (n_{i,a}(\tau)-n)( n_{j,a}(\tau')-n)( n_{k,a}(\tau'')-n) \rangle^2 \nonumber
\end{align} 
\newpage

At an instability around $Q_{ab}=0$ (which holds formally for $\tilde W=0$), a first order transition takes place if $C_{Q_{ab}^3}>C_{Tr Q^3}$. I think one can easily show that this is the case for low density, roughly because $\chi_{ij}\sim n$ and $\langle \int d\tau d\tau'd\tau''  (n_{i,a}(\tau)-n)( n_{j,a}(\tau')-n)( n_{k,a}(\tau'')-n) \rangle\sim n$.

{\bf How to treat the case of finite $\tilde W$?}
\end{comment}

\end{document}